\documentclass[%
preprint,
superscriptaddress,
bibnotes,
amsmath,amssymb,
aps,
prb,
]{revtex4-2}

\usepackage[utf8]{inputenc}
\usepackage{graphicx}
\usepackage{dcolumn}
\usepackage{bm}
\usepackage{caption}
\usepackage{subcaption}
\usepackage{multirow}
\usepackage{booktabs} 
\usepackage[normalem]{ulem}
\usepackage{hyperref}
\usepackage[usenames,dvipsnames]{xcolor}
\usepackage{tikz}
\usepackage{dsfont}
\usetikzlibrary{automata, arrows.meta, calc, fit, positioning,shapes}

\newcommand{\be}{\begin{equation}}
\newcommand{\ee}{\end{equation}}
\newcommand{\bea}{\begin{eqnarray}}
\newcommand{\eea}{\end{eqnarray}}

\newcommand{\setmuskip}[2]{#1=#2\relax}
\newcommand{\ttimes}{\ensuremath{\setmuskip{\medmuskip}{0 mu}\times}}
\usepackage{graphicx}
\usepackage{amsmath,amssymb}
\usepackage[normalem]{ulem}
\begin{document}


\title[ ]{Density functional Bogoliubov-de Gennes theory for superconductors implemented in the \textsc{SIESTA} code}

\author{R. Reho$^1$, N. Wittemeier$^2$, A. H. Kole$^1$, P. Ordejón$^2$, Z. Zanolli}

\address{Chemistry Department, Debye Institute for Nanomaterials Science and  ETSF, Condensed Matter and Interfaces, Utrecht University, PO Box 80.000, 3508 TA Utrecht, The Netherlands.}

\address{Catalan Institute of Nanoscience and Nanotechnology - ICN2 (CSIC and BIST), Campus UAB, Bellaterra, 08193 Barcelona, Spain}

\begin{abstract}
We present \textsc{SIESTA}-BdG, an implementation of the
simultaneous solution of the Bogoliubov-de Gennes (BdG) and Density Functional Theory (DFT) problem 
in  \textsc{SIESTA}, a first-principles method and code for material simulations which uses pseudopotentials and a localized basis set. 
This unified approach describes both conventional and unconventional superconducting states, and enables a description of inhomogeneous superconductors and heterostructures.
We demonstrate the validity, accuracy, and efficiency of \textsc{SIESTA}-BdG by computing physically relevant quantities (superconducting charge density, band structure, superconducting gap features, density of states) for conventional singlet (Nb, Pb) and unconventional (FeSe) superconductors.  
We find excellent agreement with experiments and results obtained within the KKR-BdG computational framework.
\textsc{SIESTA}-BdG forms the basis for modelling quantum transport in superconducting devices and including - in an approximate fashion - the superconducting DFT potential of Oliveira, Gross, and Kohn.
\end{abstract}

\maketitle

\section{Introduction}\label{sec:Introduction}
Superconductivity is an intrinsically quantum phenomenon, defying the classical picture of dissipative current flow. It is one of the striking cases where quantum effects manifest at the macroscopic scale, as demonstrated by the Meissner effect~\cite{meissner1933neuer}.
The microscopic theory developed by Bardeen, Cooper, and Schrieffer (BCS) in 1957~\cite{bardeen1957theory,bardeen1957microscopic} was instrumental in understanding the underlying mechanism in terms of attractive interaction between electrons mediated by crystal lattice vibrations (phonons), forming Cooper pairs below a (very low) critical temperature. However, there are other complex realizations of superconductivity which cannot be explained within the BCS theory. These are referred to as {\it unconventional} or exotic forms of superconductivity.
Examples include high-temperature superconductivity and multiband superconductivity, where multiple bands cross the Fermi surface, leading to several superconducting instabilities~\cite{SteglichSuperconductivityPresence1979, JeromeSuperconductivitySynthetic1980, TsueiPairingSymmetry2000, StewartSuperconductivityIron2011}. 
These and other exotic forms of superconductivity can arise due to strong interactions between electrons~\cite{balents2020superconductivity}, from the interplay between disorder and dimensionality~\cite{sacepe2020quantum}, or emerge from the proximity interaction of a conventional superconductor and a topological insulator (topological superconductivity)~\cite{frolov2020topological}.
In particular, topological superconductors~\cite{KitaevUnpairedMajorana2001, HeMagneticField2018, LeeTransportStudies2019, PradaAndreevMajorana2020} could revolutionize the field of quantum computing by hosting quantum bits based on Majorana zero modes~\cite{sato2017topological}.  

Modelling and understanding the origins of effects like proximity-induced superconductivity and unconventional superconductivity requires covering one or more of the following points: 
(a) a realistic description of heterostructures and interfaces between normal and superconducting materials (for example, to model topological superconductors)
(b) a unified treatment of normal and superconducting electronic properties 
to predict band structure, density of states, normal and anomalous charge density,  
and (c) the ability to model unconventional pairing mechanisms between electrons (e.g. $p-$wave spin-triplet), beyond BCS theory.
Superconducting density functional theory (SCDFT)~\cite{OliveiraDensityFunctionalTheory1988} satisfies these requirements: 
it combines density functional theory (DFT)~\cite{hohenberg1964inhomogeneous, kohn1965self}, which successfully predicts structural and electronic properties of solids, with the Bogoliubov-de Gennes (BdG) formalism 
~\cite{bogoljubov1958new, BogoljubovNewMethod1958, GennesSuperconductivityMetals1966},
which is a Hartree-Fock-like mean-field approximation of the many-body Hamiltonian generalizing the BCS theory~\cite{bardeen1957theory} to treat inhomogeneous superconductors. 
SCDFT allows for a general form of the pairing interaction and, hence, can describe 
normal/superconductor interfaces
and unconventional superconductors~\cite{TemmermanGapAnisotropy1996, GyorffyQuasiparticleSpectra1998}, in addition to conventional ones ~\cite{CudazzoInitioDescription2008, BoeriViewpointRoad2019, BoeriUnderstandingNovel2020}.

In SCDFT, a non-local superconductive effective potential (depending on both normal and anomalous charge densities) replaces the standard DFT one (Section~\ref{subsec:bdglcao} and Appendix~\ref{subsec:bdgdft}).
Then, the energy functional's minimization with respect to both normal and anomalous densities leads to the BdG equations.
SCDFT is formally an exact theory. However, it does not give a prescription for the construction of the superconducting coupling (kernel) $\lambda$. 
In this work, we follow the semi-phenomenological approach of Suvasini \textit{et al}~\cite{SuvasiniMultipleScattering1992, SuvasiniComputationalAspects1993}, where the superconducting kernel is assumed to be {\it local} and the explicit functional dependence on the normal ($\rho$) and anomalous ($\chi$) densities is replaced by {\it site-dependent constants} (Section~\ref{subsec:bdglcao}), namely we neglect the dynamical update of the superconducting kernel due to quasiparticle interactions (such as magnons or phonons).

Semi-phenomenological SCDFT has been recently implemented within multiple scattering theory in the Korringa-Kohn-Rostoker (KKR) approximation (KKR-BdG)~\cite{csire2015multiple, russmann_density_2022} 
to study conventional superconductors~\cite{russmann_density_2022}, proximity-induced superconductivity~\cite{russmann2022proximity, zhang2020proximity, li2023proximity}, and the formation of Yu-Shiba-Rusinov states (YSR)~\cite{nyari2021relativistic, park2023effects, laszloffy2023classification}. 
A different approach, combining dynamical mean field theory (DMFT) and the quasiparticle self-consistent GW approximation, has been recently developed and implemented in the Questaal code~\cite{pashov2020questaal} and successfully used to model (un)conventional superconductors~\cite{pashov2020questaal,acharya2022role,acharya2023vertex}.
In Questaal, the superconducting properties are extracted from pairing susceptibilities in the particle-particle channel, computed with the Bethe-Salpeter equation. 

In this paper, we present the theoretical description and software implementation of semi-empirical SCDFT in the \textsc{SIESTA} DFT method and code~\cite{SolerSIESTAMethod2002, garcia2020siesta}, which we will refer to as \textsc{SIESTA}-BdG.
\textsc{SIESTA} is a well-documented, open source code with an exhaustive set of tutorials~\footnote{https://github.com/siesta-project and https://www.youtube.com/@siestaproject}, and a large community of developers and users.
\textsc{SIESTA} uses a basis of strictly localized orbitals, which leads to sparse matrices which can be diagonalized using highly efficient linear algebra algorithms: \textsc{SIESTA} has a native interface for the ScaLAPACK, MRRR and ELPA solvers. These solvers and recently developed GPU-accelerated interface in ELPA can also be used in \textsc{SIESTA}-BdG.
In addition, the ELSI library allows simultaneous parallelization over k-points, orbitals, and spins, which is particularly useful given the extremely dense sampling of the Brillouin Zone required to capture the (sub)meV energy scales characterizing the superconducting gap. 
All these features make \textsc{SIESTA}-BdG an ideal choice for studying superconductivity in systems consisting of several hundreds of atoms such as interfaces, or heterostructures.

In \textsc{SIESTA}-BdG, the specific value of the superconducting coupling can be taken from experiment, simulations (electron-phonon DFT, strong correlation model, ...), or chosen to explore the properties of the system as a function of the input parameters.
Then, one can perform a self-consistent calculation of both normal and superconducting ({\it anomalous}) charge densities to compute the pairing potential and superconducting properties, such as local density of states (directly comparable with scanning tunnelling microscopy experiments), shape and size of the superconducting gap, quasiparticle spectrum, electron and hole wave functions, Fermi surface, and electron momentum distribution function 
(Section~\ref{subsec:solmethod}).
\textsc{SIESTA}-BdG can be employed to analyse the role that different atomic orbitals play in the formation of Cooper pairs by selectively “activating” the pairing between specific orbitals. This feature is useful to investigate the mechanism behind exotic forms of superconductivity.
\textsc{SIESTA}-BdG is implemented using the Spin-Orbit coupling (SOC) formalism present in \textsc{SIESTA}~\cite{cuadrado2012fully}, so that SOC and related effects (such as topology and topological superconductivity) can be fully described in synergy with normal state properties.
Hence, within \textsc{SIESTA}, one can compute the full spectrum of structural, electronic, magnetic, and topological properties of both normal and superconducting materials. 

Quantum electron transport can also be computed using the same software package, ensuring a coherent description of all these properties for the whole simulation workflow.
The \textsc{SIESTA}-BdG method is an ideal starting point for superconducting quantum transport calculations using non-equilibrium Green's functions techniques (NEGF), because the strictly localized basis sets used in \textsc{SIESTA} make it easy to define local Hamiltonians and local Green’s functions. Indeed, the \textsc{SIESTA} package comes with a set of routines and utilities that implement methods to compute quantum transport properties in normal state device, namely \textsc{TranSiesta}~\cite{BrandbygeDensityfunctionalMethod2002, papior2017improvements}. 
In this work, we show how the \textsc{SIESTA}-BdG method can be applied to study superconducting quantum transport in a (5,5) armchair carbon nanotube (CNT) using NEGF methods in a one-particle framework~\cite{ness2022supercurrent, ness2023ab, LambertGeneralizedLandauer1991, TaddeiSuppressionGiant1999, StefanucciTimedependentQuantum2010}.
We compute the normal and \textit{Andreev} transmission and reflection functions for an artificial N-S junction.

The manuscript is organized as follows. In Section~\ref{sec:SIESTA-BdG} we adapt the Bogoliubov-de Gennes formalism ~\cite{BogoljubovNewMethod1958} to linear combination of atomic orbitals (LCAO) methods, present the \textsc{SIESTA}-BdG implementation, three different self-consistency schemes to solve the \textsc{SIESTA}-BdG equations, and two methods for the initialization of the superconducting pairing potential and coupling.
In Section~\ref{sec:results} we present \textsc{SIESTA}-BdG simulations for conventional (Nb) and unconventional superconductors (FeSe), and validate them against experimental and/or theoretical simulations with KKR-BdG. 
In Section~\ref{sec:transport}, we formulate electron transport in the presence of superconductivity within the \textsc{SIESTA}-BdG method  and discuss the foreseen impact on the field.
In the appendices we discuss the spin generalized superconducting DFT formalism (Appendix~\ref{subsec:bdgdft}), computational strategies to perform \textsc{SIESTA}-BdG simulations (Appendix~\ref{subsec:siestabdginpractice}) including efficiency, accuracy, \textsc{SIESTA}-BdG solution methods, initialization of the superconducting potential, and computational details.
Adaptation of  scattering theory for transport in superconductors to the \textsc{SIESTA}-BdG formalism is covered in Appendix~\ref{app:transport}.

\section{\textsc{SIESTA}-BdG}\label{sec:SIESTA-BdG}

\subsection{Bogoliubov-de Gennes equations in a localized basis set}
\label{subsec:bdglcao}
In the \textsc{SIESTA}-BdG method, the spin generalized Kohn-Sham BdG equations
 - reviewed in Appendix~\ref{subsec:bdgdft} -
are written in terms of the \textsc{SIESTA} linear combination of atomic orbitals (LCAO)  basis set~\cite{SolerSIESTAMethod2002}.
An atomic orbital $\phi_{\mu}$ is identified by
an atom label ($I$) and its quantum numbers ($l, m, n$), collected in a super-index $\mu = \{I, l, m, n\}$ which labels 
the basis orbitals in the unit cell:  
\be\label{eq:LCAOorbital}
\varphi_{\mu}^{\alpha}(\mathbf{r})=\varphi_{I l m n \alpha}(\mathbf{r})=\varphi_{I l m n}(\mathbf{r}) \otimes|\alpha\rangle
\ee
with $\alpha$ the spin index.
The electron $u_i^{\alpha}(\mathbf{r})$ and hole $v_i^{\alpha}(\mathbf{r})$ eigenfunctions in the LCAO basis set can still be  grouped into four-component Nambu spinors $\Psi_{i}(\mathbf{r})=({u}_i^{\uparrow}\left(\mathbf{r}\right) u_i^{\downarrow}\left(\mathbf{r}\right) v_i^{\uparrow}\left(\mathbf{r}\right) v_i^{\downarrow}\left(\mathbf{r}\right))^T$,
and read:
\be 
 u_i^\alpha(\mathbf{r})=\sum_\mu \varphi_\mu(\mathbf{r}) u_{i\mu}^{\alpha}~, \quad    
 v_i^\alpha(\mathbf{r})=\sum_\mu \varphi_\mu^*(\mathbf{r}) v_{i\mu}^{\alpha} 
\ee
where $u_{i\mu}^{\alpha}$ and $v_{i\mu}^{\alpha}$ are the expansion coefficients, and $i$ the index of the superconducting bands (Bogoliubon excitation state).

\textsc{SIESTA} solves the DFT problem using periodic boundary conditions.
Non-periodic systems (molecules, tubes, slabs, interfaces, ...) are modelled using  the supercell approach. 
Assuming a lattice periodic pairing potential, it is convenient to write and solve the BdG eigenvalue problem in the reciprocal space of the simulation cell. Exploiting the Bloch theorem, the electron and hole wavefunction can be expanded in Bloch waves
\be 
    u^{\alpha}_{i}(\mathbf{k},\mathbf{r})
    ={}  \sum_{\mu, \mathbf{R}} e^{i\mathbf{k}(\mathbf{R}+\boldsymbol{\tau_\mu})} \varphi_\mu(\mathbf{r}) u_{i \mu}^{\alpha}(\mathbf{k})~,  \label{eq:bdg-uk-local}
    \quad 
    v^{\alpha}_{i}(\mathbf{k}, \mathbf{r})
    ={}  \sum_{\mu, \mathbf{R}} e^{i\mathbf{k}(\mathbf{R}+\boldsymbol{\tau_\mu})} \varphi^*_\mu(\mathbf{r}) v_{i \mu }^{\alpha}(\mathbf{k})
\ee
where the sum runs through all orbitals in space, given by the lattice vectors $\mathbf{R}$ and orbitals positions $\boldsymbol{\tau}_{\mu}$ in the unit cell, and $\mathbf{k}$ is the crystal momentum of the supercell.
Writing the BdG equations in the LCAO basis is crucial to reveal the contribution of specific orbitals to superconductivity, explain complex interactions between different bands (multi-band superconductors),  
and analyse competing superconducting phases as those breaking time-reversal symmetry or related to different spin configurations~\cite{WangMagneticGround2016}. 
The relevant equations for the \textsc{SIESTA}-BdG method in $\mathbf{k}$-space are:
\begin{enumerate} 
\item the BdG Hamiltonian $\mathbf{\widehat{H}_{BdG}}$

\be\label{eq:HSpinBdGkspaceLCAO}
\widehat{H}_{BdG, \mu\nu}^{\alpha,\beta}(\mathbf{k}) = \sum_{\nu, \mathbf{R}} e^{i\mathbf{k}(\mathbf{R}+\boldsymbol\tau_{\nu}-\boldsymbol{\tau}_{\mu})} \widehat{H}^{\alpha,\beta}_{BdG,\mu\nu}
\ee
\item the eigenvalue problem
\bea \label{eq:SpinBdGkspaceLCAO} \nonumber
&\sum_{\mu\alpha}\left(\begin{array}{cc}
        \mathbf{h^{KS}}^{\beta\alpha}_{\nu\mu}(\mathbf{k}) -\mu S_{\nu\mu}(\mathbf{k})\delta_{\beta\alpha}
      & \Delta^{\beta\alpha}_{\nu\mu}(\mathbf{k})
    \\ 
       -\Delta^{*\beta\alpha}_{\nu\mu}(-\mathbf{k})
      & -\mathbf{h^{KS}}^{*\beta\alpha}_{\nu\mu}(-\mathbf{k}) + \mu S_{\nu\mu}(\mathbf{k})\delta_{\beta\alpha}
    \end{array}\right)\left(\begin{array}{c}
         u^{\alpha}_{i\mu}(\mathbf{k})  \\
         v_{i\mu}^{\alpha} (\mathbf{k})
    \end{array}\right) \\ 
    & = \varepsilon_i \sum_{\mu\alpha}S_{\nu\mu}(\mathbf{k})\delta_{\beta\alpha}\left(\begin{array}{c}
         u^{\alpha}_{i\mu}(\mathbf{k})  \\
         v_{i \mu}^{\alpha} (\mathbf{k})
    \end{array}\right),
\eea
where $\bf{h^{KS}}$ is the normal state Kohn-Sham Hamiltonian, $\Delta^{\beta\alpha}_{\nu\mu}(\mathbf{k})$ is the superconducting pairing potential, $S_{\nu\mu}(\mathbf{k})$ the overlap matrix at $\mathbf{k}$ and $\mu$ the chemical potential,
\item the normal and anomalous densities
\bea\label{eq:densitieslcaokspace} 
\rho^{\alpha \beta}_{\mu\nu}(\mathbf{k}) 
    = \sum_i f\left(\varepsilon_i(\mathbf{k})\right)u_{i\mu}^ {*\alpha}(\mathbf{k}) u_{i\nu}^{\beta}(\mathbf{k}) + 
      \sum_i \bar{f}(\varepsilon_i(\mathbf{-k})) v_{i \mu}^{\alpha}(\mathbf{-k})v_{i \nu}^{*\beta}(\mathbf{-k}) \\
\chi^{\alpha\beta}_{\mu\nu} (\mathbf{k})
   =\sum_i f\left(\varepsilon_i(\mathbf{k})\right) u_{i\nu}^{\beta}(\mathbf{k})v_{n\mu}^{*\alpha}(\mathbf{k}) + 
    \sum_i \bar{f}\left(\varepsilon_i(\mathbf{-k})\right) u_{i\mu}^{\alpha}(\mathbf{-k})v_{i\nu}^{*\beta}(\mathbf{-k})
\eea
where $f(\varepsilon_i)$ and $\bar{f}(\varepsilon_i)=1-f(\varepsilon_i)$ are the occupation functions,

\item the pairing potential $\Delta^{\alpha\beta}_{\mu\nu}(\mathbf{k})$:
\begin{equation}\label{eq:dlambdachi}
\Delta^{\alpha\beta}_{\mu\nu}(\mathbf{k}) = -\sum_{l=0,1,2,3}\sum_{\gamma} i \sigma_l^{\alpha\gamma} \sigma_2^{\gamma\beta}
    \int_{\mathrm{BZ}} \mathrm{d}\mathbf{k^\prime}\lambda_{l}^{\mu\nu}(\mathbf{k}-\mathbf{k^\prime})\chi^{l}_{\mu\nu}(\mathbf{k^\prime})
\end{equation}
where $\sigma_l$ are the Pauli matrices, $\lambda_{l}^{\mu\nu}$ are the orbital projected superconducting couplings. We denote as $\boldsymbol{\lambda}$ the matrix with elements $\lambda_{l}^{\mu\nu}$.
\end{enumerate}

The above equations are only valid under the assumption that the pairing potential is lattice periodic. Superconductors with spatially non-uniform pairing potential need to be modelled using a supercell approach. This is the case of Fulde-Ferrell-Larkin-Ovchinnikov (FFLO) phase of a superconductor~\cite{casalbuoni2004inhomogeneous}, which is characterized by Cooper pairs with finite momentum.

Even though the superconducting coupling $\boldsymbol{\lambda}$ is local,
it can have an explicit dependence on the atomic orbitals in real-space, which  
can be used to initialize the simulations (Sections~\ref{subsec:specifypairingpot} and \ref{sec:bulkPb}). 
The pairing potential $\Delta$ is updated in the self-consistent steps via~\ref{eq:dlambdachi} (Section~\ref{subsec:solmethod}) and it acquires a non-trivial wave vector $\mathbf{k}$-dependence.
Hence, the \textsc{SIESTA}-BdG formalism allows one to compute non-isotropic superconducting band gap and superconductors with multiple $\bf{k}$-dependent gaps (multiband superconductors).

In the following, the quantities computed in real and reciprocal space are denoted with a $\mathbf{R}$ or $\mathbf{k}$ superscript, respectively (e.g. $\mathbf{H}^{\mathbf{R}}$,  $\mathbf{H}^{\mathbf{k}}$). 
The matrix elements of any function in real space $g(\mathbf{r})$ can be written in the LCAO basis as 
$g^{\mathbf{R}}_{\mu,\nu}
        = \sum_{\mathbf{R}'} \int_{\text{UC}+\mathbf{R}'} \mathrm{d^3}\mathbf{r} \phi^*_\mu(\mathbf{r}) g(\mathbf{r}) \phi_\nu(\mathbf{r}-\mathbf{R})$.

\subsection{\textsc{SIESTA}-BdG SCF Solution Methods}\label{subsec:solmethod} 
Preliminary to \textsc{SIESTA}-BdG simulations, we perform a 
normal state DFT calculation to compute the electronic Hamiltonian $\bf{h^{KS}}$ and normal state density $\rho$. 
The converged $\bf{h^{KS}}$ is used to construct the BdG Hamiltonian $\mathbf{H}_{BdG,\mu\nu}^\mathbf{R}$ by introducing a pairing potential $\boldsymbol{\Delta}$ to break charge conservation and enable the superconducting phase.
In \textsc{SIESTA}-BdG, $\boldsymbol{\Delta}$ and $\boldsymbol{\lambda}$ can be initialized in the orbital or real space representation, as discussed in Section \ref{subsec:specifypairingpot}. 
Then, the \textsc{SIESTA}-BdG Hamiltonian and (normal and anomalous) densities can be computed by solving the eigenvalue problem (Eq.~\ref{eq:SpinBdGkspaceLCAO}) either directly (\textit{non SCF-BdG}) or by performing a self-consistent field cycle (SCF) on only the normal density (\textit{fixed-$\Delta$}) or both the normal and anomalous densities (\textit{full SCF-BdG}).

For a given choice of $\boldsymbol{\Delta}$ or $\boldsymbol{\lambda}$ full self-consistency should also include the chemical potential in addition to the densities. The number of electrons is given by a weighted sum over all states, using a (chemical potential dependent) occupation function as the weighting factor. For normal-state calculations, only the occupation function depends on the chemical potential ($\mu$) while for \textsc{SIESTA}-BdG calculations, $\mu$ explicitly enters the eigenvalue problem. This means that a change in $\mu$ requires one to solve the eigenvalue problem again to find new solutions, making the process more complex. 
To compute $\mu$ we follow the approach by Suvasini et al.~\cite{SuvasiniComputationalAspects1993} where $\mu$ is updated in an outer loop that repeats the SCF cycle until the correct number of particles is reached~\cite{SuvasiniComputationalAspects1993}. 
The user can skip this loop to reduce computational costs. In this case, the total number of electrons might not match the number of electrons in the normal state, but experience shows that the deviation is small for typical values of the pairing potential~\cite{SuvasiniComputationalAspects1993}.
The \textsc{SIESTA}-BdG solution methods are illustrated in Figure~\ref{fig:scdft_scheme} and described below.

\begin{figure}[h]
    \centering
    \scalebox{0.62}{
    \begin{tikzpicture}[font=\small,align=center,node distance=0.75cm]
        \definecolor{borderColour}{HTML}{2f528f}
        \definecolor{myBlue}{HTML}{4472c4}
        \definecolor{myOrange}{HTML}{ed7d31}
        \definecolor{myGreen}{HTML}{92d050}
        
        \tikzset{box/.style={
                draw=borderColour, very thick,
                minimum width=2.5cm,
                inner sep=2mm,
                align=center,
                minimum height=1cm}}
        \tikzset{input/.style={
                trapezium,
                trapezium left angle = 65,
                trapezium right angle = 115,
                trapezium stretches}}
        \tikzset{choice/.style={
                diamond,
                aspect=2.5}}
        \tikzset{connect/.style={->, >=latex, ultra thick, borderColour}}

        \node[box,rounded rectangle] (START) at (0, 0) {START};
        \node[box,input,below=of START] (Rho0) {DFT SCF: Find self-consistent solution for the \\
            normal-state electronic density
            $\rho_0$};  

        \node[box,below=1cm of Rho0] (HofRho) {Construct $h^{\mathrm{KS}}_{\mu\nu}[\rho_0]$};
        \node[box,choice,below=1.5cm of HofRho ] (Oneshot) {non SCF-BdG\\ method?};          
                
        \node[box,below=1.2cm of Oneshot] (Diag) {Solve eigenvalue problem: \\
            $\left(\begin{array}{cc}
                h^{KS}_{\mu\nu} & \Delta_{\mu\nu}\\
                -\Delta^*_{\mu\nu} & -h^{KS}*_{\mu\nu}
            \end{array}\right)
            \left(\begin{array}{c}
                u_{\nu}\\
                v_{\nu}
            \end{array} \right)
            = \varepsilon \left(\begin{array}{cc}
                \mathbf{S}_{\mu\nu} & 0\\
                0 & \mathbf{S}_{\mu\nu}
            \end{array} \right)
            \left(\begin{array}{c}
                u_{\nu}\\
                v_{\nu}
            \end{array} \right)$};
        \node[box,below=of Diag] (CalcRhoChiFixD0) {Construct $\rho$ and $\chi$ matrices
            from $\left(\begin{array}{c} u_{\nu}\\ v_{\nu}\end{array} \right)$};            
          \node[box,choice,below=of CalcRhoChiFixD0 ] (FixDorFixL) {Fixed-$\Delta$ or \\ full SCF-BdG?};
        \node[box,rounded rectangle,left=2cm of FixDorFixL] (FixD) {Fixed-$\Delta$};
        \node[box,below=1.2cm of FixD] (CalcHk) {Construct $h^{KS}_{\mu\nu}$};
        \node[box,choice,below=of CalcHk ] (SCFixD) {Self-consistent\\
        $\rho$ and $h^{KS}_{\mu\nu}$?};        
        \node[box,rounded rectangle,right=2cm of FixDorFixL] (FixL) {full SCF-BdG};
        \node[box,below=1.2cm of FixL] (CalchDFixL) {Construct $h^{KS}_{\mu\nu}$ and $\Delta_{\mu\nu}$};
        \node[box,choice,below=of CalchDFixL] (SCFixL) {Self-consistent\\
        $\rho$, $h^{KS}_{\mu\nu}$, $\chi$, $\Delta$?}; 
        \node[box,choice,below=5cm of FixDorFixL] (Charge) {Charge conserved?}; 
        \node[box,choice,below=of Charge] (Fermi) {Update Fermi level?}; 
        \node[box,below=1.5cm of Fermi] (Postprocess) {Solve eigenvalue problem: \\
            $\left(\begin{array}{cc}
                h^{KS}_{\mu\nu} & \Delta_{\mu\nu}\\
                -\Delta^*_{\mu\nu} & -h^{KS}*_{\mu\nu}
            \end{array}\right)
            \left(\begin{array}{c}
                u_{\nu}\\
                v_{\nu}
            \end{array} \right)
            = \varepsilon \left(\begin{array}{cc}
                \mathbf{S}_{\mu\nu} & 0\\
                0 & \mathbf{S}_{\mu\nu}
            \end{array} \right)
            \left(\begin{array}{c}
                u_{\nu}\\
                v_{\nu}
            \end{array} \right)$ \\
            for bands, PDOS, $\chi$, etc.};        
        
        \draw[connect] (START) -- (Rho0);
        \draw[connect] (Rho0) -- (HofRho);
        \draw[connect] (HofRho) -- (Oneshot) node[pos=0.5,fill=white,inner sep=1mm]{Input $\Delta_{\mu\nu}$};
        \draw[connect] (Oneshot) -- ++(13cm,0)   node[pos=0.5,fill=white,inner sep=1mm]{Yes}  |-node[pos=0.25,fill=white,inner sep=1mm]{\textit{skip} \\ \textsc{SIESTA}-BdG\\\textit{SCF loop}} (Postprocess);
        \draw[connect] (Oneshot) -- (Diag) node[pos=0.35,fill=white,inner sep=1mm]{No};
        \node (SCF) [box,draw=myOrange,rounded corners,inner sep=11mm,fit = (Diag) (SCFixD) (SCFixL)(Fermi),label={[anchor=center,fill=white,font=\large,text=myOrange]135:{\textbf{\textsc{SIESTA}-BdG SCF loop}}},
        minimum width=21.3cm] {};        
        \draw[connect] (Diag) -- (CalcRhoChiFixD0);   
        \draw[connect] (CalcRhoChiFixD0) -- (FixDorFixL);
        \draw[connect] (FixDorFixL) -- (FixD);
        \draw[connect] (FixDorFixL) -- (FixL);
        \draw[connect] (FixD) -- (CalcHk);
        \draw[connect] (CalcHk)-- (SCFixD);
        \draw[connect] (SCFixD) -- (Charge) node[pos=0.35,fill=white,inner sep=1mm]{Yes};
        \draw[connect] (SCFixD) -- ++ (-3.91cm,0) node[pos=0.5,fill=white,inner sep=1mm]{No} |- (Diag);
        \draw[connect] (FixL) -- (CalchDFixL)node[pos=0.5,fill=white,inner sep=1mm]{Input $\lambda$};
        \draw[connect] (CalchDFixL) -- (SCFixL);
        \draw[connect] (SCFixL) -- (Charge)
        node[pos=0.35,fill=white,inner sep=1mm]{Yes};
        \draw[connect] (SCFixL) -- ++ (4.275cm,0) node[pos=0.5,fill=white,inner sep=1mm]{No} |- (Diag);
        \draw[connect] (Charge) -- (Fermi)
        node[pos=0.35,fill=white,inner sep=1mm]{No};
        \draw[connect] (Charge) -- ++(7cm,0) node[pos=0.5,fill=white,inner sep=1mm]{Yes} |- (Postprocess);
        \draw[connect] (Fermi) -- ++(-10.250cm,0) node[pos=0.5,fill=white,inner sep=1mm]{Yes} |- (Diag);
        \draw[connect] (Fermi) -- (Postprocess) node[pos=0.5,fill=white,inner sep=1mm]{No};
    \end{tikzpicture}}
    \caption{Schematic of the iterative process to compute self-consistent solutions of the \textsc{SIESTA}-BdG secular equation. During the update step within the \textsc{SIESTA}-BdG SCF loop the density or the Hamiltonian is mixed.}
    \label{fig:scdft_scheme}
\end{figure}

\textbf{Non SCF-BdG}: 
The $\mathbf{H}_{BdG}^\mathbf{R}$ Hamiltonian is constructed and diagonalized (Eq.~\ref{eq:HSpinBdGkspaceLCAO}). No \textsc{SIESTA}-BdG SCF loop is performed; therefore, $\bf{h^{KS}}$ is not updated in the presence of $\boldsymbol{\Delta}$. 
The resulting eigenvalues and eigenvectors are post-processed to calculate electronic band structure, density of states, normal and anomalous charge density. 
The underlying assumption is that superconductivity is a small perturbation of $\bf{h^{KS}}$ and the normal state density.     
The \textit{non SCF-BdG} method has a significantly lower computational cost compared to \textit{fixed-$\Delta$} and \textit{full SCF-BdG}, as the latter two methods perform an additional SCF cycle using the four-component Nambu spinors $\Psi_i(\mathbf{r})$. 

\textbf{Fixed-$\bf{\Delta}$}: 
The $\mathbf{H}_{BdG}^\mathbf{R}$ Hamiltonian is diagonalized in the presence of a fixed pairing potential $\boldsymbol{\Delta}$. At every step of the \textsc{SIESTA}-BdG SCF cycle, the Kohn-Sham Hamiltonian ($\bf{h^{KS}}$), the normal ($\rho$) and the anomalous ($\chi$) densities are updated. Self-consistency is ensured on the normal-state part ($\bf{h^{KS}}$, $\rho$) of the \textsc{SIESTA}-BdG Hamiltonian.
The underlying assumption is that 
the perturbations induced on the superconducting properties due to the anomalous charge are negligible, namely
the initial guess for $\boldsymbol{\Delta}$ is good enough to model the superconducting system and does not need to be updated in the SCF cycle via Eq.~\ref{eq:dlambdachi}.
In particular, for bulk Nb the difference between \textit{fixed-$\Delta$} and \textit{non SCF-BdG} in the density of states are negligible for values of the pairing potential $\mathbf{\Delta}$ smaller than $\sim$ 100 meV (Appendix~\ref{subsec:comparingnonscfbdgfixeddelta}).

\textbf{full SCF-BdG}: is an implementation of the full self-consistent field \textsc{SIESTA}-BdG scheme.  The superconducting couplings $\boldsymbol{\lambda}$ 
and the external pairing potential $d^i_{\mathrm{ext}}(\mathbf{r},\mathbf{r^\prime})$ (Eq.~\ref{eq:dlocal}) must be specified in the input. 
The external pairing potential is needed to ensure a non-zero anomalous density in the first SCF step, and is set to zero thereafter (see discussion after Eq.~\ref{eq:dlocal}). At each SCF step, $\bf{h^{KS}}$, $\rho$, $\chi$ and $\boldsymbol{\Delta} = \sum_{i} \left(d_{\text {ext }}^i\left(\mathbf{r}\right) -\lambda^{i}\left(\mathbf{r}\right) \chi^i\left(\mathbf{r}\right)\right)  \sigma_i i \sigma_2$ are updated.
Analogously to normal-state simulations, $\boldsymbol{\Delta}$ and $\chi$ can be mixed using generalized \textsc{\textsc{SIESTA}} algorithms.
Convergence can be achieved on $\boldsymbol{\Delta}$, on $\chi$, or on both of them simultaneously.\label{enum:Fixed-lambda}\\

\subsection{Initialization of the pairing potential and superconducting coupling}\label{subsec:specifypairingpot}

\begin{figure}
    \centering
    \includegraphics[width=\linewidth]{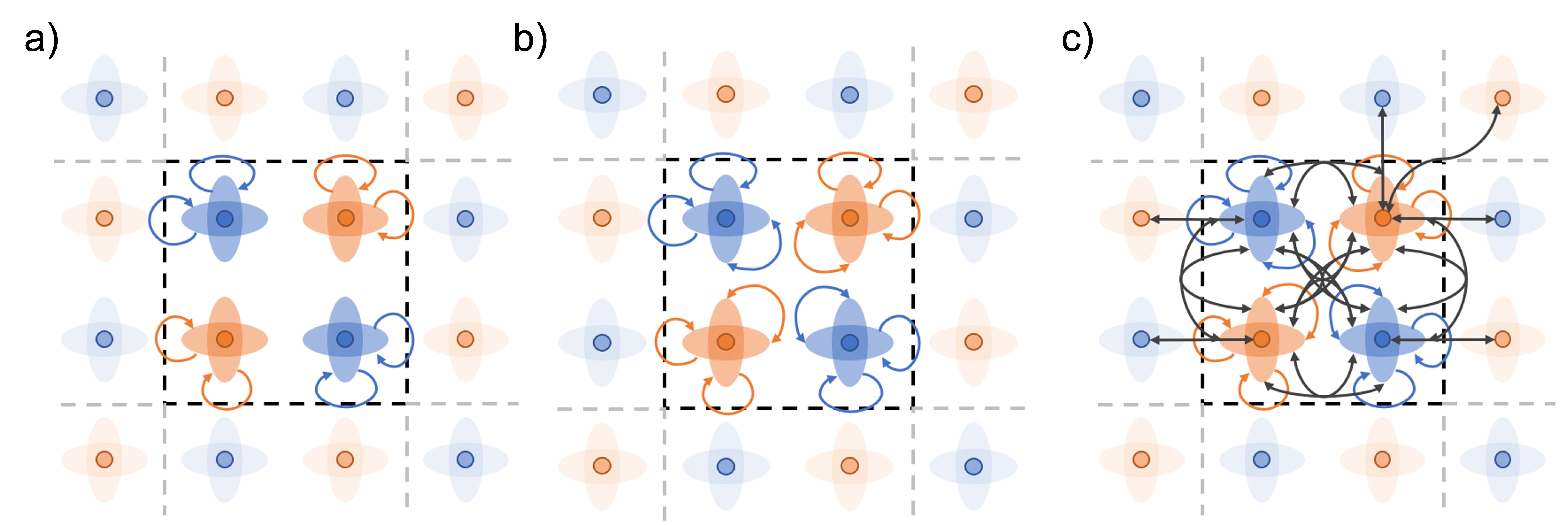}
    \caption{Initialization of the superconducting pairing in the \textit{orbital representation}: Schematic representation of coupling between orbitals in the (a) intra-orbital, (b) on-site, and (c) full overlap pairing.}
    \label{fig:pairing-types}
\end{figure}

Depending on the solution method, either the superconducting coupling $\boldsymbol{\lambda}$ (\textit{full SCF-BdG}) or the pairing potential $\boldsymbol{\Delta}$ ({\it{non SCF-BdG}} and {\it{fixed-$\Delta$}}) is a semi-empirical parameter. 
Their initial values can be expressed either in terms of the LCAO atomic orbitals (\textit{orbital representation})
or as a modulation in real space of a superconducting strength parameter $\bar{\lambda}$ or $\bar{\Delta}$ (\textit{superconducting strength representation}) by a function $g(\mathbf{r})$, i.e. $\Delta(\mathbf{r}) = \bar{\boldsymbol{\Delta}}g(\mathbf{r})$.
Note that, when working in this representation, we use a bold symbol to indicate real-space matrices with orbital projected matrix elements (as $\boldsymbol{\lambda}$) and italics for real space functions $\lambda(\mathbf{r})$.
The \textit{orbital representation} is ideal for investigating the nature of the superconducting interaction by selectively coupling specific orbitals. 
The user can directly specify the matrix elements of the pairing potential ($d^{i \mathbf{R}}_{\mu\nu}$)  for any superconducting channel ($i$). The full pairing potential matrix is constructed using the orbital representation of Eq.~\ref{eq:deltalocal}:
\bea
    \boldsymbol{\Delta}^{\mathbf{R}}_{\mu\nu} = \sum_{i=0,1,2,3} d^{i \mathbf{R}}_{\mu\nu}  \sigma_i i \sigma_2 
    = \sum_{i=0,1,2,3} \left(d_{\text {ext},\mu\nu}^{i\mathbf{R}} -\lambda^{\mathbf{R}\mu\nu}_{i}\chi^{i\mathbf{R}}_{\mu\nu}\right)  \sigma_i i \sigma_2
\eea
If the symmetries of the gap function are known, then the initial guess for $\mathbf{\Delta}$ can be constructed as a matrix respecting those symmetries. For instance, one can specify 
a pairing exclusively between electrons and holes occupying the same orbital on the same atom (\textit{intra-orbital}, diagonal $\boldsymbol{\Delta}$), or between different electron and hole orbitals belonging to the same atom (\textit{on-site}, block-diagonal $\boldsymbol{\Delta}$), or couple all the orbitals of all atoms (\textit{full overlap}, full $\boldsymbol{\Delta}$  matrix), as illustrated in Figure~\ref{fig:pairing-types}. The flexibility of the \textit{orbital representation} has been used to identify the orbitals responsible for unconventional superconductivity in FeSe (Section~\ref{subsec:fese}).

The \textit{superconducting strength representation} has a more intuitive physical meaning, as the superconducting pairing is defined on the \textsc{SIESTA} real-space grid. In this case, the user can choose among the different options depending on the knowledge of the physical properties and symmetries of the materials to be studied.
The computed physical observables (for instance, $\chi$) mainly depends on the spatial basis overlap $S_{\mu\nu}^\mathbf{R}$ and not on $g(\mathbf{r})$, as illustrated in Appendix~\ref{sec:bulkPb} for bulk Pb.
The current implementation includes the following options for the modulation function $g(\mathbf{r})$:
\begin{itemize}
    \item
    \textit{Gaussians}: $g(\mathbf{r};\mathbf{x},\sigma) = e^{-|\mathbf{r}-\mathbf{x}|^2/(2\sigma^2)}$, with center $\mathbf{x}$ and width $\sigma$. 
    \item
    \textit{Spherical wells}: $g(\mathbf{r};\mathbf{x},\sigma) = \theta(|\mathbf{r}-\mathbf{x}|;\sigma,\eta)$ with center $\mathbf{x}$ and radius $\sigma$.  The shape $\theta$ can be a Heaviside step function (hard well) or a Fermi-Dirac function with broadening $\eta$ (soft well).
    \item
    \textit{Plane waves}: $g(\mathbf{r};\mathbf{k},\phi) = \cos(\mathbf{k}\mathbf{r}+\phi)$, where $\mathbf{k}$ is the wave vector and $\phi$ is the phase of the wave at $\mathbf{x}=0$.
    \item any combination of the above
\end{itemize}
Gaussian and spherical wells can be centred at any point within the unit cell and have arbitrary widths, potentially extending beyond the unit cell boundary. The lattice periodicity is automatically imposed. 
Plane waves can be used to represent fully delocalized interactions. Superpositions of multiple waves can be used to create pairing potentials with different symmetries.
The versatility of the initial choice can be used to model complex systems with inhomogeneous superconducting potential. 

\section{Results}\label{sec:results}

\subsection{Bulk Niobium}\label{sec:Nb}

\begin{figure}
    \centering
    \includegraphics[width=0.85\textwidth]{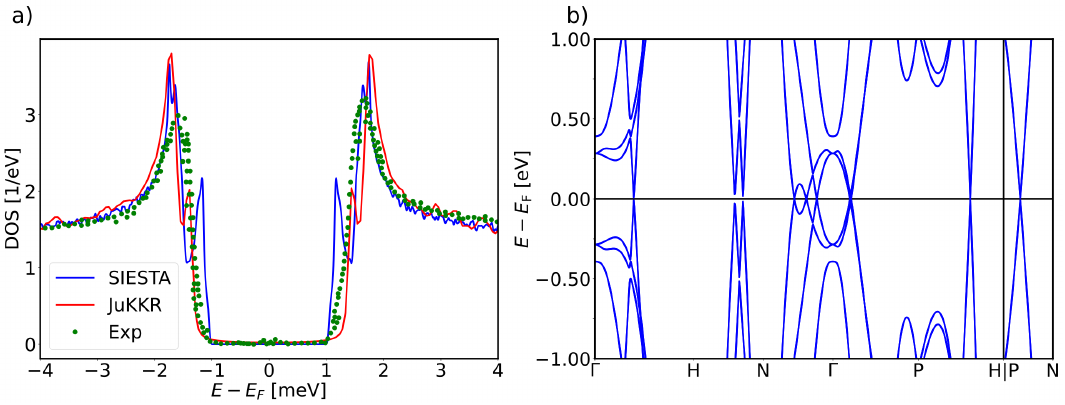}
        \caption{Bulk Nb: (a) Density of States (DOS) comparison among \textit{full SCF-BdG} \textsc{SIESTA},  JuKKR ~\cite{russmann_density_2022, philipp_density_2021} 
        and STM measurements~\cite{schneider_realizing_nodate}, showing good agreement between simulations and experiment. 
        For \textsc{SIESTA}-BdG we set $\lambda= 0.188$~eV$\cdot V_\mathrm{sphere}$ with a spherical ($r = 1.4136$ \AA) square well potential and a smearing temperature $T = 0.1$ K.
    In the JuKKR code a value of $\lambda = 1.11$~eV was used~\cite{russmann_density_2022}. 
    (b) Superconducting band structure with visible particle-hole symmetry. 
}
    
    \label{fig:NbDOScomparison2}
\end{figure}

Bulk Niobium is a good system for validating the \textsc{SIESTA}-BdG method, as several theoretical~\cite{SuvasiniComputationalAspects1993, russmann_density_2022} and experimental~\cite{PhysRev.149.231, karasik1970superconducting,schneider_realizing_nodate} results are available in the literature for comparison.
We computed the superconducting state DOS using the \textit{full SCF-BdG} method and compared the \textsc{SIESTA}-BdG result with the JuKKR code~\cite{russmann_density_2022} and experimental data from STM measurements~\cite{schneider_realizing_nodate} (Figure~\ref{fig:NbDOScomparison2}). There is good agreement between the two codes: both predict a DOS with two peaks at~$\sim\pm2$ meV (main coherence peaks), characteristic of conventional superconductors, and denoted U-shaped gap.
In addition, the two codes predict visible gap anisotropy, as revealed by the multiple peaks in the DOS for low enough smearing temperatures. The two approaches differ in the position of the innermost peaks, the relative shift is about $\pm 0.2$~meV, which can be explained by differences in the assumptions underlying the two codes.
The overall gap shape also matches the one from the experiment. The missing gap anisotropy in the experimental data could be due to temperature (smearing), 
\textsc{SIESTA}-BdG predicts an additional
gap anisotropy (i.e. each main coherence peak splits in two) compared to JuKKR, possibly due to the finer sampling of the energy axis in our calculation.

\subsection{Bulk FeSe}\label{subsec:fese}

\begin{figure*}
    \centering
    \includegraphics[width=\textwidth]{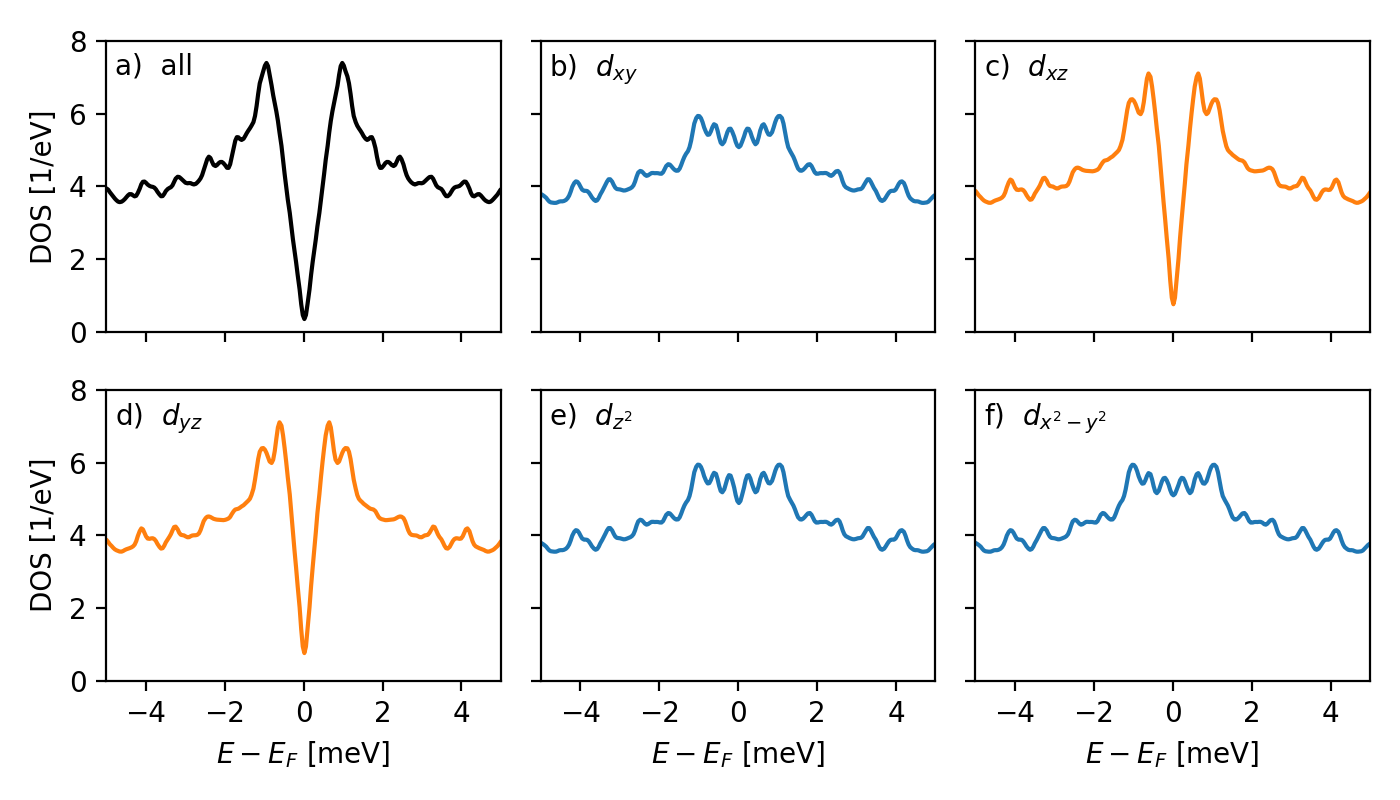}
    \caption{DOS of FeSe, an unconventional superconductor, for different pairing potentials calculated with the \textsc{SIESTA}-BdG method: (a) isotropic pairing ($\Delta(\mathbf{r})$ = 3~meV), (b-f) pairing potential limited to a different Fe-$d$ orbital ($\Delta_{d_\nu d_\nu}=3$~meV). The simulations with isotropic pairing (a)  reproduce the size and shape of the experimentally observed superconducting gap~\cite{KasaharaFieldinducedSuperconducting2014}. The superconducting gap emerges only if the pairing potential is applied to the $d_{xz}$ (c) or $d_{yz}$ (d)
     orbitals.}
    \label{fig:dos_FeSe}
\end{figure*}

The origin and character of high-temperature superconductivity in FeSe (and other iron-based superconductors) are currently debated~\cite{GlasbrennerEffectMagnetic2015, song2011direct, hirschfeld2016using}. It has been suggested that FeSe is an unconventional $d$-wave superconductor~\cite{PaglioneHightemperatureSuperconductivity2010, BaekOrbitaldrivenNematicity2015, MassatChargeinducedNematicity2016} driven by pairing interaction in the $3d_{xz}$ and $3d_{yz}$ orbitals of iron~\cite{ZhangObservationTwo2015}. 
Experimental evidence suggests that FeSe does not possess long-range magnetic order. Even at low temperatures (4~K), spin fluctuations are significant~\cite{WangMagneticGround2016}. 
These fluctuations occur primarily between two magnetic phases: Néel (or checkerboard) and staggered dimer. In the N{\'e}el magnetic structure, the magnetic moments of neighbouring iron atoms are antiparallel. In the following, we report results for the Néel magnetic structure.
The simulation of the nematic phase would require very large supercells in periodic boundary conditions, which is beyond the scope of this work.

The superconducting gap of FeSe in the N{\'e}el phase 
is V shaped and 
of the order of $\sim 1-10$ meV  at the coherence peaks, depending on the experimental conditions~\cite{KasaharaFieldinducedSuperconducting2014}.
To model the superconducting state we use 
the \textit{non SCF-BdG} method and initialize the pairing potential with a constant ($\Delta(\mathbf{r}) = \bar\Delta=3$~meV). 
The correct V shape of the superconducting gap emerges from our calculations (Figure~\ref{fig:dos_FeSe}.a). This indicates that orbital symmetries are sufficient to predict the conventional or unconventional nature of the superconducting gap.

To analyse the role of different orbitals in enabling superconductivity, we restrict the pairing interaction to specific orbitals:
We use the \textit{orbital representation} method (Section~\ref{subsec:specifypairingpot}) and assign non-zero values to matrix elements corresponding to a specific Fe-$3d$ orbital (Fe-$3d_{xz}$, Fe-$3d_{yz}$, \dots) at a time. 
We observe that imposing a pairing potential exclusively between $3d_{xz}$ or $3d_{yz}$ orbitals 
opens a V-shaped gap (Figure~\ref{fig:dos_FeSe}~b-f).
We can therefore identify those orbitals as responsible for unconventional superconductivity in FeSe.
The superconducting gap is smaller than the one computed with isotropic pairing among {\it all} orbitals because the pairing interaction is limited to a subset of all orbitals, reducing the pairing potential's average strength. 
No superconducting gap emerges if the pairing interaction is restricted to any of the other Fe-$3d$ orbitals.

\section{Quantum Transport within the \textsc{SIESTA}-BdG formalism}
\label{sec:transport}

\subsection{The formalism}
The SCDFT formalism maps the many-body problem, including the pair interaction giving rise to Cooper pairs, onto an equivalent non-interacting problem described by one-particle wave functions.
In this auxiliary system, many aspects of quantum transport can be described in one-particle frameworks in a Landauer-based conductivity approach. As is commonly done in DFT approaches, the KS wavefunctions are assumed to be a good approximation to the one-particle wavefunction of the interacting many-body system~\cite{NessInitioTransport2023,LambertGeneralizedLandauer1991,TaddeiSuppressionGiant1999,StefanucciTimedependentQuantum2010}. However, any current described in such a one-particle framework is fundamentally a one-particle current. As such, Cooper pair currents are not described. 

In a non-superconducting device, the current through any electrode $\mathfrak{e}$ at temperature $T_\mathfrak{e}$ and chemical potential $\mu_\mathfrak{e}$ is given by~\cite{DattaElectronicTransport1995}
\begin{align}
     I_\mathfrak{e} 
     = \frac{e}{h}
     \int_{-\infty}^\infty\hspace{-0.8em}d\varepsilon\,
     \int_{\mathrm{BZ}}\hspace{-0.6em}d\mathbf{k}\,
       \sum_{\mathfrak{e}'\neq\mathfrak{e}} T_{\mathfrak{e}\mathfrak{e}'} [f_\mathfrak{e}(\varepsilon) - f_{\mathfrak{e}'}(\varepsilon)]       
\end{align}
where $T_{\mathfrak{e}\mathfrak{e}'}$ is the transmission function between two electrodes $\mathfrak{e}$ and $\mathfrak{e}'$, and $f_{\mathfrak{e}}(\varepsilon)$ is the occupation function for the electrode $\mathfrak{e}$.
The energy and $\mathbf{k}$ dependence of the quantities is implicit.

In the case of superconducting electrodes or devices, the single-particle current can be carried either by the electrons or holes. 
The transmission ($T_{\mathfrak{e}\mathfrak{e}'}$) and reflection probabilities ($R_{\mathfrak{e}}$) become 2x2 matrices
\begin{align}\label{eq:RTscmatrices}
    R_\mathfrak{e} &\rightarrow \begin{pmatrix}
        R_\mathfrak{e}^{pp} & R_\mathfrak{e}^{hp} \\ 
        R_\mathfrak{e}^{ph} & R_\mathfrak{e}^{hh}  
    \end{pmatrix} &
    T_{\mathfrak{e}\mathfrak{e}'} &\rightarrow \begin{pmatrix}
       T_{\mathfrak{e}\mathfrak{e}'}^{pp} & T_{\mathfrak{e}\mathfrak{e}'}^{hp}\\
        T_{\mathfrak{e}\mathfrak{e}'}^{ph} & T_{\mathfrak{e}\mathfrak{e}'}^{hh}  
    \end{pmatrix}
\end{align}
with $R_\mathfrak{e}^{pp}$ ($R_\mathfrak{e}^{hh}$) the reflection probability for electrons (holes) in lead $\mathfrak{e}$ to scatter back into electronic (hole) states on the same electrode, and $R_\mathfrak{e}^{ph}$ ($R_\mathfrak{e}^{hp}$) the reflection probability for an electron (hole) in lead $\mathfrak{e}$ to scatter into hole (electron) states in the same lead (Andreev reflection). 
Similarly, the four components of $T_{\mathfrak{e}\mathfrak{e}'}$ describe the scattering probabilities between electron and hole states in lead $\mathfrak{e}$ and electron and hole states in $\mathfrak{e}'$ (normal and anomalous transmission). 
The transmission and reflection functions can be calculated using the Green's function formulation of quantum transport~\cite{DattaElectronicTransport1995, FerrerGOLLUMNextgeneration2014a}.
Using all four functions (see Appendix~\ref{app:transport} for additional details), the current through any (normal) electrode can be written as~\cite{LambertPhasecoherentTransport1998}
\begin{align}\label{eq:currentsc}
    I_{\mathfrak{e}} = \frac{2e}{h}
     \int_{-\infty}^\infty\hspace{-0.8em}d\varepsilon\,
     \int_{\mathrm{BZ}}\hspace{-0.6em}d\mathbf{k}\,
     \Big\{
            &\sum_{\mathfrak{e}'\neq\mathfrak{e}} 
                T^{pp}_{\mathfrak{e}\mathfrak{e}'} [
                    f_{\mathfrak{e}}(\varepsilon) - f_{\mathfrak{e}'}(\varepsilon)
                  ]
                - T^{hh}_{\mathfrak{e}\mathfrak{e}'} [
                    \bar{f}_{\mathfrak{e}}(\varepsilon) - \bar{f}_{\mathfrak{e}'}(\varepsilon)
                  ]
            \nonumber\\ 
            +&\sum_{\mathfrak{e}'\neq\mathfrak{e}} 
                T^{ph}_{\mathfrak{e}\mathfrak{e}'} 
                [f_{\mathfrak{e}}(\varepsilon) 
                -\bar{f}_{\mathfrak{e}'}(\varepsilon)]
                -T^{hp}_{\mathfrak{e}\mathfrak{e}'} 
                [\bar{f}_{\mathfrak{e}}(\varepsilon)
                -f_{\mathfrak{e}'}(\varepsilon)]
            \nonumber\\
            +&[R^{hp}_\mathfrak{e} 
            + R^{ph}_\mathfrak{e}] [f_{\mathfrak{e}}(\varepsilon) - \bar{f}_{\mathfrak{e}}(\varepsilon)]
        \Big\}
\end{align}
where $\bar{f}_{\mathfrak{e}}(\varepsilon)$ is the occupation of hole states in the electrode $\mathfrak{e}$. Moving to the low-bias limit $ f_\mathfrak{e}(\varepsilon) 
    = f(\varepsilon) - \frac{\partial{f}}{\partial{\varepsilon}}(\varepsilon) [\mu_\mathfrak{e}-\mu]$ ($\mu$ the reference chemical potential w.r.t to which the electrode potentials are shifted) and invoking particle-hole symmetry, the expression for the current can be rewritten as:
\begin{align}\label{eq:currentsclowbias}
    I_{\mathfrak{e}} 
        \approx& \frac{2e}{h}\int_{\mathrm{BZ}}\hspace{-0.6em}d\mathbf{k}\,\sum_{\mathfrak{e}'\neq\mathfrak{e}}
            T^{pp}_{\mathfrak{e}\mathfrak{e}'}(\mu) [\mu_{\mathfrak{e}'} - \mu_{\mathfrak{e}}]
        +\sum_{\mathfrak{e}'\neq\mathfrak{e}} 
            T^{ph}_{\mathfrak{e}\mathfrak{e}'}(\mu) [2\mu - \mu_{\mathfrak{e}'} - \mu_{\mathfrak{e}}]
        +2 R^{hp}_\mathfrak{e}(\mu) [\mu - \mu_\mathfrak{e}]
\end{align}

In this formula three different components of the current can be identified: 
(i) a \textit{normal} component which depends on electron-electron (hole-hole) transmission ($T^{pp}$, $T^{hh}$), 
(ii) an \textit{anomalous} component which depends on electron-hole (hole-electron) transmission ($T^{ph}$, $T^{hp}$), 
and (iii) an \textit{Andreev} contribution which depends solely on the Andreev reflections ($R^{hp}$, $R^{ph}$).
Particle conservation requires that the currents flowing in all electrodes sum up to zero (Kirchhoff's current law~\cite{KirchhoffUeberDurchgang1845}): $\sum_\mathfrak{e}I_{\mathfrak{e}}=0$~\cite{LambertGeneralizedLandauer1991,datta1996scattering}. 
This condition uniquely determines the chemical potential $\mu$: 
The current is not proportional to the pairwise differences in the electrode chemical potentials, and explicitly depends on the 
electrode chemical potentials. 
This is an important difference with respect to a normal-state device. 

Eq.~\ref{eq:currentsc} is valid for computing the current of any normal electrode with a superconducting device. However, it is not suitable for a superconducting electrode $\mathfrak{e}$, as it does not account for the contributions of Cooper pair currents. Arguments on how to properly compute currents under non-equilibrium conditions and in superconducting electrodes are given in~\cite{StefanucciTimedependentQuantum2010, LambertGeneralizedLandauer1991,datta1996scattering} and references therein.

A non-zero conductance at zero applied voltage (zero-bias conductance) can only arise due to Andreev (or anomalous) terms in this formalism. In a bulk superconductor, no reflections occur, and the anomalous transmission is equal to zero. Thus, the zero-bias conductance in any bulk superconductor is zero. Only in the presence of a scattering region can Andreev reflections (or anomalous transmission) occur and lead to a non-zero conductance. 
This, of course, does not imply a finite resistance of the bulk superconductor, since Cooper pair currents are not included in the description. 

The \textsc{SIESTA}-BdG method, due to the strictly localized basis set, constitute an ideal starting point for quantum transport calculations using non-equilibrium Green's functions techniques (NEGF) on open quantum systems with superconducting components, such as superconductor junctions and topological superconductor devices.
The chemical potential in the leads does not change as a consequence of the interaction with the scattering region~\cite{DattaElectronicTransport1995}.

\begin{figure}[h]
    \includegraphics[width=\textwidth]{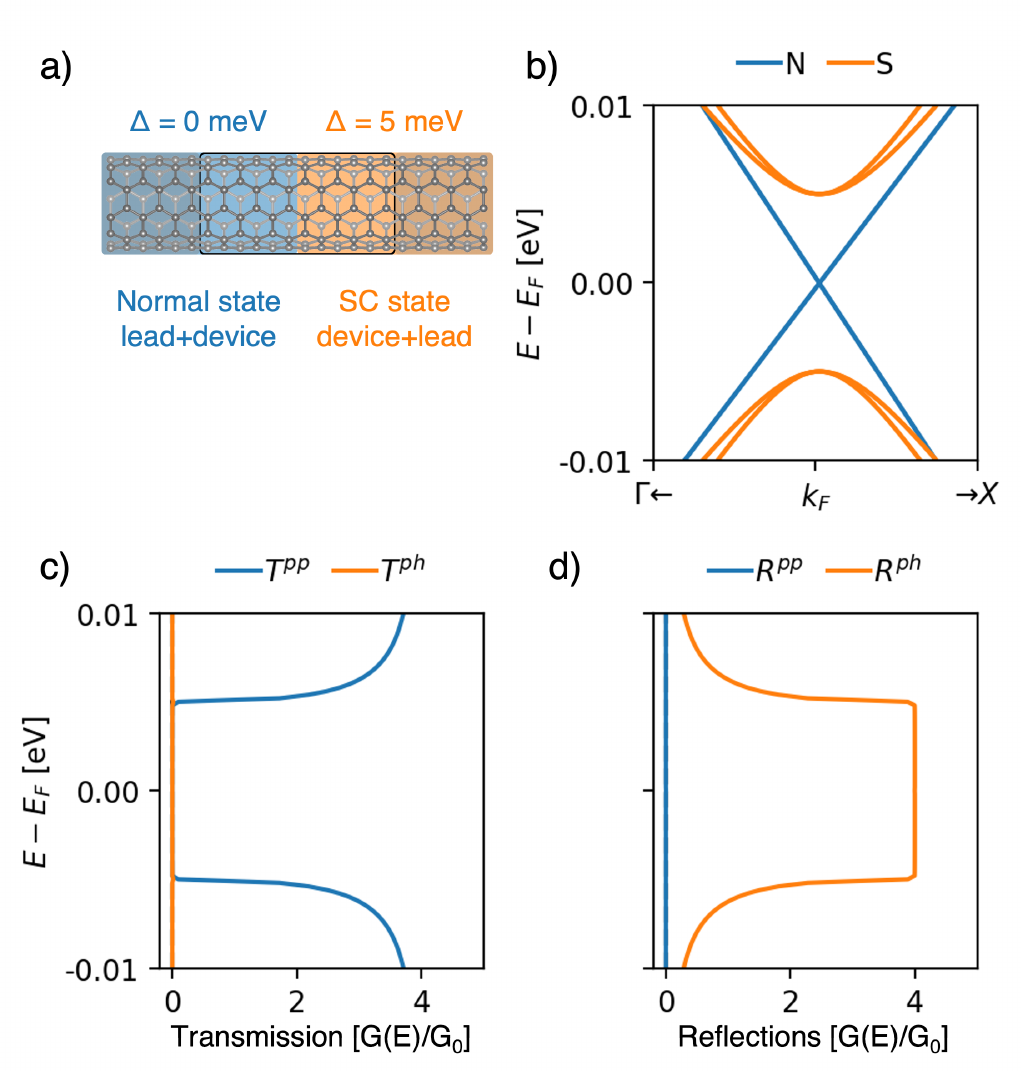}
    \caption{
    Normal-Superconductor (N-S) junction in an (5,5) carbon nanotube.
    The normal ($\Delta=0$ meV) and superconducting ($\Delta=5$ meV) regions are indicated in blue and orange, respectively. 
    (a) Simulation cell for transport simulations displaying scattering region and leads (shaded area) 
    (b) Electronic band structure of the scattering region of the N-S  (5, 5) carbon nanotube junction.
    (c) Transmission and (d) reflection  in quantum conductance units $G_0$. Within the superconductor gap, all electrons in the normal electrode participate in Andreev reflections ($R^{eh}$). 
    For energies larger than the superconducting gap, the N-S junction behaves like a normal metal. At the superconducting gap there is a smooth transition between the two regimes. 
}
    \label{fig:cnt}
\end{figure}

To showcase the possible application of the \textsc{SIESTA}-BdG method in the quantum transport formalism, we consider a (5,5) armchair carbon nanotube (CNT) as a transport system.
Signatures of superconductivity have been observed in different types of carbon nanotubes: (a) 4~\AA~  single-wall carbon nanotubes at $T_C\approx 15$~K (Meissner effect)~\cite{TangSuperconductivityAngstrom2001}, (b) multi-wall nanotubes with 10~nm outer tube at $T_C\approx 12$~K (sudden drop of DOS below $T_C$)~\cite{TakesueSuperconductivityEntirely2006}, and (c) ropes of single-wall carbon nanotubes at 0.55~K (drop of resistance)~\cite{KociakSuperconductivityRopes2001}. However, it is difficult to reliably create superconductivity in one-dimensional nanostructures due to low DOS and quantum fluctuations~\cite{SasakiTheorySuperconductivity2007}. If realized and stabilized, superconducting carbon nanotubes could be ideal one-dimensional leads for nanoelectronics. 

We construct an N-S junction within a (5,5) nanotube by selectively applying the pairing potential to part of the scattering region and right electrode. Our transport device consists of 3 rings of the (5,5) nanotube (scattering region) which are contacted by the same nanotube on each side (electrodes, each 3 rings wide), thus forming an infinite tube. We use the Hamiltonian obtained from a periodic calculation neglecting the coupling with the infinite leads. We then use the Green's function formulation of quantum transport~\cite{DattaElectronicTransport1995} to obtain the transmission and reflection probabilities.
We apply the \textit{non SCF-BdG} method and neglect any changes in the electronic structure of the leads  due to the superconducting pairing (Landauer assumption). 
Therefore, the resulting N-S junction possesses a negligible barrier for the electrons and electron-electron back-scattering should be close to non-existent. Thus, our N-S junction is expected to behave like an ideal N-S junction, allowing for  Andreev reflections that can lead to non-vanishing reflection coefficients. Furthermore, we use an isotropic superconducting pairing with a strength of 5~meV.

In the normal state, the band structure of a (5,5) armchair CNT near the Fermi level (blue iti7lines in Figure~\ref{fig:cnt}.a) is described by four linear bands (two per spin channel). In the superconducting state (orange lines, Figure~\ref{fig:cnt}.a), a band gap opens up at the Fermi level. The band dispersion becomes parabolic, and the number of bands is doubled. 
In addition, Figure~\ref{fig:cnt} shows the different transmission and reflection functions for the N-S junction between a normal state CNT and superconducting CNT.
The Andreev reflections $R^{ph}$ and normal transmission $T^{pp}$ (Figures~\ref{fig:cnt}.b and~\ref{fig:cnt}.c) become significant at the edge of the superconducting gap where electron and hole states hybridize in the scattering region because of the pairing potential. 
Deep within the gap, the anomalous transmission $T^{ph}$ and the normal reflection $R^{pp}$ are zero, and all incoming states are perfectly reflected (Figures~\ref{fig:cnt}.b and~\ref{fig:cnt}.d). These results exactly replicate the expected behaviour for a barrier-free N-S junction~\cite{BlonderTransitionMetallic1982}.

\section*{Conclusions and Outlook} \label{sec:conclusion}
In this work we present \textsc{SIESTA}-BdG, a method to calculate the superconducting properties of materials and its implementation in the open-source \textsc{SIESTA}  code ~\cite{garcia2020siesta}.
\textsc{SIESTA}-BdG implements  the semi-phenomenological SCDFT formalism of Suvasini~\cite{SuvasiniComputationalAspects1993} 
in  localized basis sets 
(equations~\ref{eq:SpinBdGkspaceLCAO} and~\ref{eq:dlambdachi}).
We show that \textsc{SIESTA}-BdG  can accurately predict both conventional  and unconventional superconductors. 

We propose three  solution methods to solve the DFT-BdG secular eqution~\ref{eq:SpinBdGkspaceLCAO}: \textit{non SCF-BdG}, \textit{fixed-$\Delta$} and \textit{full SCF-BdG}, which mainly differ in the treatment of self-consistency of the superconducting density and Hamiltonian. We present different options for initializing the superconducting coupling/pairing, which is a parameter in this method.
We sustained our discussion with three test cases:  Nb, Pb and FeSe bulk, which demonstrate the robustness and accuracy of our implementation. We critically compared our results with experiments and KKR-BdG simulations, finding excellent quantitative agreement (Section~\ref{sec:results}). 
We introduced and discussed optimization schemes such as the adaptive Brillouin Zone sampling, the tetrahedron method, parallelization over $\mathbf{k}$-points and  real space initialization of the pairing potential that are indispensable for an accurate description of the superconducting state.
We conclude the manuscript with a perspective on the simulation of quantum transport for superconductors 
by combining the \textsc{SIESTA}-BdG  with the NEGF formalism  
of Ref.~\cite{BrandbygeDensityfunctionalMethod2002, papior2017improvements}. As a proof of concept, we compute transmission and reflection probabilities of a normal/superconducting junction in a carbon nanotube.

The limitations of \textsc{SIESTA}-BdG   are related to 
the ignorance of the exact form of the superconducting coupling ($\boldsymbol{\lambda}$) or pairing ($\boldsymbol{\Delta}$), namely the magnitude and phase of the superconducting interaction is unknown. 
One can make an estimate using full SCDFT for conventional superconductors, 
or refer to theories that model the superconducting coupling strength itself~\cite{in2023screening, simonato2023revised}.
As a guideline to initialize the pairing potential/coupling, we suggest using first the \textit{superconducting strength representation} with a constant modulation function to acquire familiarity with the system and, then, test more specific real-space initializations.
When the symmetries of the system are known, it might be easier to impose them directly in the matrix elements of the pairing potential ({\it orbital representation}, Section~\ref{subsec:solmethod}).
Exploring the consequences of the initialization of the superconducting parameter (different solution methods, new or more complex real space initialization, etc.) needs to be addressed case-by-case, as it is material specific.

The applicability of our approach is vast, heterostructures and complex device configurations being prime candidates that could benefit from this novel implementation, given \textsc{SIESTA}'s inherent optimal scaling. 
In perspective, \textsc{SIESTA}-BdG framework can be interfaced with existing SCDFT codes, mapping the non-local SCDFT potential into an approximate local one
and/or extended to a non-local superconducting potential to allow for SCDFT calculations.
\textsc{SIESTA}-BdG is the basis for the description of transport in superconducting systems:
In general, the SCDFT formalism maps the many-body problem, including the pair interaction giving rise to Cooper pairs, onto an equivalent non-interacting problem described by one-particle wavefunctions.
In this auxiliary system, transport is described by a one-particle framework (e.g. scattering matrix formalism or Green's function formalisms~\cite{stefanucci2010time}), as opposed to other Many-Body formalism which focuses on the two-particle superconducting response functions~\cite{acharya2022role,acharya2023vertex,acharya2021electronic}.
The current implementation will provide a useful tool to further investigate and understand exotic forms of superconductivity and related emerging physics such as unconventional superconductivity, proximity induced superconductivity, topological superconductivity, interplay between magnetism and superconductivity.

\section*{Acknowledgments}
The authors acknowledge the fruitful discussion with Philipp R\"u{\ss}mann and Andr\'es R. Botello-Mend\'ez, Miguel Pruneda, and Gábor Csire, Antonio Sanna, Herv\'e Ness, and Enrico Perfetto.
ZZ acknowledges the research program “Materials for the Quantum Age” (QuMat) for financial support. This program (registration number 024.005.006) is part of the Gravitation program financed by the Dutch Ministry of Education, Culture and Science (OCW).
RR and AK financial support from Sector Plan Program 2019-2023. 
This work was sponsored by NWO-Domain Science for the use of supercomputer facilities. We also acknowledge that the results of this research have been achieved using the Tier-0 PRACE
Research Infrastructure resource Discoverer based in Sofia, Bulgaria  (OptoSpin project id. 2020225411). 
This project has received funding from the European Union’s Horizon Europe research and innovation programme under grant agreement No 101130384 (QUONDENSATE).
PO and NW acknowledge support from the EU MaX CoE (Grant No. 101093374) and grants PCI2022-134972-2 and PID2022-139776NB-C62 funded by the Spanish MCIN/AEI/10.13039/501100011033 and by the ERDF, A way of making Europe, and Grant No.  
NW further acknowledges funding from  the  European  Union’s  Horizon  2020  research  and  innovation  programme  under  the  Marie  Skłodowska-Curie  Grant  Agreement  No.  754558  (PREBIST  –  COFUND).  
ICN2 is supported by the Severo Ochoa programme from Spanish MINECO (grant no. CEX2021-001214-S) and by Generalitat de Catalunya (CERCA programme).

\section*{Data availability statement}
The data that support the findings of this study are openly available on Materials Cloud. 
\appendix 

\section{Spin generalized Superconducting DFT}\label{subsec:bdgdft}
The SCDFT Hamiltonian $\widehat{H}_{\mathrm{BdG}}\left(\mathbf{r}, \mathbf{r}^{\prime}\right)$~\cite{OliveiraDensityFunctionalTheory1988}
consists of four blocks describing
particles (top-left), antiparticles (bottom-right), and particle/antiparticle coupling $\Delta$ (off-diagonal) 
\footnote{We interchangeably use the term electrons (holes) and particles (anti-particles).}: 

\bea \label{eq:KSBdGrealspace}
\widehat{H}_{\mathrm{BdG}}\left(\mathbf{r}, \mathbf{r}^{\prime}\right) = &\left(\begin{array}{cc}
        \widehat{h}(\mathbf{r})\delta(\mathbf{r}-\mathbf{r^\prime})
      & \Delta(\mathbf{r},\mathbf{r^\prime})
    \\ 
       \Delta^{*}(\mathbf{r},\mathbf{r^\prime})
      & -\widehat{h}^{*}(\mathbf{r})\delta(\mathbf{r}-\mathbf{r^\prime})
    \end{array}\right)
\eea

To extend the $2\times2$ SCFDT Hamiltonian to the $4\times4$ fully relativistic case  (i.e. including SOC), we follow the \textsc{SIESTA} formalism~\cite{garcia2020siesta}: 
\bea \label{eq:KSBdGrealspacespin}
\widehat{H}_{\mathrm{BdG},\alpha\beta}\left(\mathbf{r}, \mathbf{r}^{\prime}\right) = &\left(\begin{array}{cc}
        \widehat{h}^{KS}_{\alpha\beta}(\mathbf{r})\delta(\mathbf{r}-\mathbf{r^\prime})
      & \Delta_{\alpha\beta}(\mathbf{r},\mathbf{r^\prime})
    \\ 
       -\Delta_{\alpha\beta}^{*}(\mathbf{r},\mathbf{r^\prime})
      & -\widehat{h}^{*KS}_{\alpha\beta}(\mathbf{r})\delta(\mathbf{r}-\mathbf{r^\prime})
    \end{array}\right)
\eea
\noindent with spin indices $\alpha$ and $\beta$, and
\be
\widehat{h}^{KS}(\mathbf{r}) = \left(-\nabla^2+\widehat{V}^{H}-\mu\right) \mathds{1}_2+\widehat{V}^{\mathrm{KB}}+\widehat{V}^{\mathrm{SO}} + \widehat{V}^{XC}
\ee
the $2\times2$ fully relativistic electronic Kohn-Sham Hamiltonian. 
The latter includes the kinetic energy $\nabla^2$, the Hartree potential $\widehat{V}^{H}$, the chemical potential $\mu$, 
the Kleinman–Bylander scalar-relativistic pseudo-potential $\hat{V}^{KB}$, 
the spin-orbit term $\hat{V}^{SO}$, and the exchange-correlation potential $\widehat{V}^{XC}$.
The relativistic contributions to the pairing potential $\boldsymbol{\Delta}$ are neglected.
The minus sign in the bottom-left block of Eq.~(\ref{eq:KSBdGrealspacespin}) arises from the anticommutation rules of the spin indices: $-\Delta^*_{\alpha,\beta} = \Delta^*_{\beta,\alpha}$.
Writing explicitly the spin dependence, it reads:
\be
\widehat{H}_{BdG}(\mathbf{r},\mathbf{r'}) =  \left(\begin{array}{cccc}
\widehat{h}^{KS}_{\uparrow \uparrow}(\mathbf{r}) & \widehat{h}^{KS}_{\uparrow \downarrow}(\mathbf{r}) & \Delta_{\uparrow \uparrow}\left(\mathbf{r}, \mathbf{r}^{\prime}\right) & \Delta_{\uparrow \downarrow}\left(\mathbf{r}, \mathbf{r}^{\prime}\right) \\
\widehat{h}^{KS}_{\downarrow \uparrow}(\mathbf{r}) & \widehat{h}^{KS}_{\downarrow \downarrow}(\mathbf{r}) & \Delta_{\downarrow \uparrow}\left(\mathbf{r}, \mathbf{r}^{\prime}\right) & \Delta_{\downarrow \downarrow}\left(\mathbf{r}, \mathbf{r}^{\prime}\right) \\
-\Delta_{\uparrow \uparrow}^*\left(\mathbf{r}, \mathbf{r}^{\prime}\right) & -\Delta_{\uparrow \downarrow}^*\left(\mathbf{r}, \mathbf{r}^{\prime}\right) & -\widehat{h}^{*KS}_{\uparrow \uparrow}(\mathbf{r}) & -\widehat{h}^{*KS}_{\uparrow \downarrow}(\mathbf{r}) \\
-\Delta_{\downarrow \uparrow}^*\left(\mathbf{r}, \mathbf{r}^{\prime}\right) & -\Delta_{\downarrow \downarrow}^*\left(\mathbf{r}, \mathbf{r}^{\prime}\right) & -\widehat{h}^{*KS}_{\downarrow \uparrow}(\mathbf{r}) & -\widehat{h}^{*KS}_{\downarrow \downarrow}(\mathbf{r})
\end{array}\right)
\ee
The spin generalized Kohn-Sham BdG equations are, then: 
\be \label{eq:EOMKSBdGrealspace}
\int \mathrm{d} \mathbf{r}^{\prime} \widehat{H}_{\mathrm{BdG}}\left(\mathbf{r}, \mathbf{r}^{\prime}\right)\left(\begin{array}{c}
u_i^{\uparrow}\left(\mathbf{r}^{\prime}\right) \\
u_i^{\downarrow}\left(\mathbf{r}^{\prime}\right) \\
v_i^{\uparrow}\left(\mathbf{r}^{\prime}\right) \\
v_i^{\downarrow}\left(\mathbf{r}^{\prime}\right)
\end{array}\right)=\varepsilon_i\left(\begin{array}{c}
u_i^{\uparrow}(\mathbf{r}) \\
u_i^{\downarrow}(\mathbf{r}) \\
v_i^{\uparrow}(\mathbf{r}) \\
v_i^{\downarrow}(\mathbf{r})
\end{array}\right).
\ee
The eigenfunctions of the BdG Hamiltonian are four-component Nambu spinors $\Psi_{i}(\mathbf{r})=({u}_i^{\uparrow}\left(\mathbf{r}\right) u_i^{\downarrow}\left(\mathbf{r}\right) v_i^{\uparrow}\left(\mathbf{r}\right) v_i^{\downarrow}\left(\mathbf{r}\right))^T$, where $u^{\alpha}_i(r)$ and $v^{\alpha}_i(r)$ are the electron and hole components.
In the superconducting ground state ($\varepsilon_0$) all Cooper pairs are in the same energy level.
The eigenenergies $\varepsilon_i$ above the ground state correspond to Bogoliubons, namely quasiparticles consisting of a linear combination of excited electrons and holes.
The BdG equations~(\ref{eq:KSBdGrealspacespin}) satisfy particle-hole symmetry~\cite{sato2017topological}. Namely, if  $(u_i^{\uparrow}\left(\mathbf{r}\right) u_i^{\downarrow}\left(\mathbf{r}\right) v_i^{\uparrow}\left(\mathbf{r}\right) v_i^{\downarrow}\left(\mathbf{r}\right))^T$ is an eigenvector with solution $\varepsilon_i$, 
then $(v_i^{\uparrow}\left(\mathbf{r}\right) v_i^{\downarrow}\left(\mathbf{r}\right) u_i^{\uparrow}\left(\mathbf{r}\right) u_i^{\downarrow}\left(\mathbf{r}\right))^{*T}$ leads to an equivalent solution with eigenvalue $-\varepsilon_i$.

In analogy with Kohn-Sham DFT, one can define single particle ($V$)
and pairing ($\bf{\Delta}$) effective potentials 
in terms of a superconducting exchange-correlation functional $\Omega_{xc}[\rho, \boldsymbol{\chi}]$, which is a functional of both the normal  $\rho(\mathbf{r})$ and the anomalous $\boldsymbol{\chi}(\mathbf{r},\mathbf{r'})$ charge densities :
\bea \label{eq:effectivefields} 
V[\rho, \boldsymbol{\chi}](\mathbf{r}) & = V^{H}(\mathbf{r}) + V^{KB}(\mathbf{r}) + V^{SO}(\mathbf{r})+\frac{\delta \Omega_{x c}[\rho, \boldsymbol{\chi}]}{\delta \rho(\mathbf{r})} \\
\label{eq:extpotentials}
\boldsymbol{\Delta}[\rho,\boldsymbol{\chi}]\left(\mathbf{r}, \mathbf{r}^{\prime}\right) & = d^0\left(\mathbf{r}, \mathbf{r}^{\prime}\right) \mathrm{i} \hat{\sigma}_2 + \sum_{j=1, 2, 3} d^{j}\left(\mathbf{r}, \mathbf{r}^{\prime}\right) \hat{\sigma}_j \mathrm{i} \hat{\sigma}_2 
\label{eq:delta-channel}\\ 
d^i[\rho,\boldsymbol{\chi}]\left(\mathbf{r}, \mathbf{r}^{\prime}\right) & =d_{\mathrm{ext}}^i\left(\mathbf{r}, \mathbf{r}^{\prime}\right)-\frac{\delta \Omega_{x c}[\rho, \boldsymbol{\chi}]}{\delta \chi^{*i}\left(\mathbf{r}, \mathbf{r}^{\prime}\right)}
  ~~~\mathrm{ with}~i = 0, 1, 2, 3
\eea
\noindent
where $\hat{\sigma}_{i}$ are the Pauli matrices,
$d^i$ are  the singlet ($i = 0$) and triplet ($i = 1,2,3$) spin components in the Balian-Werthamer (BW) representations~\cite{balian1963superconductivity}, and $d^i_{\mathrm{ext}}(\mathbf{r},\mathbf{r^\prime})$ is an external pairing potential introduced, formally, to break the charge symmetry of the system. 
In practice, the \textsc{SIESTA}-BdG implementation builds the BdG Hamiltonian from  $d^i_{\mathrm{ext}}(\mathbf{r},\mathbf{r^\prime})$ in the first self-consistent field iteration. The resulting anomalous density is used to build the pairing potential in the successive step and $d^i_{\mathrm{ext}}(\mathbf{r},\mathbf{r^\prime})$ is set to zero thereafter. For this reason, $d^i_{\mathrm{ext}}(\mathbf{r},\mathbf{r^\prime})$ is missing in equation(\ref{eq:dlambdachi}).

In the BW representation, the singlet and triplet spin components of the anomalous charge density
are
\bea
& \chi^0\left(\mathbf{r}, \mathbf{r}^{\prime}\right)=\frac{\chi^{\uparrow \downarrow}\left(\mathbf{r}, \mathbf{r}^{\prime}\right)-\chi^{\downarrow \uparrow}\left(\mathbf{r}, \mathbf{r}^{\prime}\right)}{2}=\frac{\chi^{\uparrow \downarrow}\left(\mathbf{r}, \mathbf{r}^{\prime}\right)+\chi^{\uparrow \downarrow}\left(\mathbf{r}^{\prime}, \mathbf{r}\right)}{2}, \\
& \chi^{1}\left(\mathbf{r}, \mathbf{r}^{\prime}\right)=\frac{\chi^{\downarrow \downarrow}\left(\mathbf{r}, \mathbf{r}^{\prime}\right)-\chi^{\uparrow \uparrow}\left(\mathbf{r}, \mathbf{r}^{\prime}\right)}{2}=\frac{\chi^{\uparrow \uparrow}\left(\mathbf{r}^{\prime}, \mathbf{r}\right)-\chi^{\downarrow \downarrow}\left(\mathbf{r}^{\prime}, \mathbf{r}\right)}{2}, \\
& \chi^{2}\left(\mathbf{r}, \mathbf{r}^{\prime}\right)=\frac{\chi^{\downarrow \downarrow}\left(\mathbf{r}, \mathbf{r}^{\prime}\right)+\chi^{\uparrow \uparrow}\left(\mathbf{r}, \mathbf{r}^{\prime}\right)}{2 \mathrm{i}}=\frac{\chi^{\downarrow \downarrow}\left(\mathbf{r}^{\prime}, \mathbf{r}\right)+\chi^{\uparrow\uparrow}\left(\mathbf{r}^{\prime}, \mathbf{r}\right)}{-2 \mathrm{i}}, \\
& \chi^{3}\left(\mathbf{r}, \mathbf{r}^{\prime}\right)=\frac{\chi^{\uparrow \downarrow}\left(\mathbf{r}, \mathbf{r}^{\prime}\right)+\chi^{\downarrow \uparrow}\left(\mathbf{r}, \mathbf{r}^{\prime}\right)}{2}=\frac{\chi^{\uparrow \downarrow}\left(\mathbf{r}, \mathbf{r}^{\prime}\right)-\chi^{\uparrow \downarrow}\left(\mathbf{r}^{\prime}, \mathbf{r}\right)}{2}.
\eea
The BW parametrization is useful to study the properties of $\boldsymbol{\chi}$ under parity transformation $\mathbf{r}\rightarrow\mathbf{r^\prime}$.
$\boldsymbol{\chi}$ is a fermionic object and, hence, must satisfy the Pauli exclusion principle, namely it must be antisymmetric under exchange of the two electrons forming the Cooper pair: $\chi^{\beta\alpha}(\mathbf{r}^{\prime},\mathbf{r})=-\chi^{\alpha\beta}(\mathbf{r},\mathbf{r}^{\prime})$. This implies that if $\chi^{\alpha\beta}(\mathbf{r},\mathbf{r}^{\prime})$ is even (odd) under the parity transformation, then it has to be odd (even) under spin exchange leading to a $S = 0$ singlet state ($S = 1$, triplet state). 

The normal and anomalous densities that minimize the energy functional $E[\rho,\chi]$ are~\cite{OliveiraDensityFunctionalTheory1988}
\be \label{eq:scdfdensities}
\begin{gathered}
\rho(\mathbf{r})=\sum_{\alpha, i} f\left(\varepsilon_i\right)\left|u_i^\alpha(\mathbf{r})\right|^2+
\sum_{\alpha, i} \bar{f}\left(\varepsilon_i\right)\left|v_i^\alpha(\mathbf{r})\right|^2 
\\
\chi^{\alpha \beta}\left(\mathbf{r}, \mathbf{r}^{\prime}\right)=\sum_i f\left(\varepsilon_i\right) u_i^\beta\left(\mathbf{r}^{\prime}\right) v_i^{*\alpha}(\mathbf{r}) 
+\sum_i \bar{f}\left(\varepsilon_i\right) u_i^\alpha(\mathbf{r}) v_i^{*\beta}\left(\mathbf{r}^{\prime}\right)
\end{gathered}
\ee
where $f(\varepsilon_i)$ is the occupation function of the $i^{th}$ Bogoliubon, and $\bar{f}(\varepsilon_i) = 1-f(\varepsilon_i)$. 
In contrast to conventional DFT, the sum over $\alpha$, and $i$ in Eq.~\ref{eq:scdfdensities} extends over all states and is not limited by the Fermi level. Furthermore, the contribution of the $uv^*$ terms in $\chi$ is expected to be non-zero only in an energy region close to the Fermi surface~\cite{GennesSuperconductivityMetals1966}. 

In SCDFT, $\Omega_{xc}[\rho,\boldsymbol{\chi}]$  
is written in terms of a superconducting kernel which is, in general, a non-local pairing interaction dependent on four spatial (and spins) variables:
$\lambda [\rho, \boldsymbol{\chi}] \left(\mathbf{r}, \mathbf{r}^{\prime} ; \mathbf{r_1}, \mathbf{r_1}^{\prime}\right)$.
In this work, we use Suvasini~\cite{SuvasiniComputationalAspects1993} assumption of a local superconducting kernel $\lambda(\mathbf{r})$. 
As a consequence, the exchange-correlation functional reduces to 

\be \label{eq:SCOmegaxc}
 \Omega_{x c}[\rho,\boldsymbol{\chi}] = \Omega_{0,xc}[\rho] + \int \sum_{i} \mathrm{d}\mathbf{r} \chi^{i*}(\mathbf{r})\lambda^i(\mathbf{r}) \chi^i(\mathbf{r})
\ee

with $\Omega_{0,xc}$  the Kohn-Sham exchange-correlation potential, and 
\be \label{eq:dlocal}
d^i\left(\mathbf{r}\right) = d_{\text{ext }}^i\left(\mathbf{r}\right) 
 -\lambda^{i}\left(\mathbf{r}\right) \chi^i\left(\mathbf{r}\right)
\ee
which is expanded in its tensor form in the BW representation as
\be \label{eq:deltalocal}
    \boldsymbol{\Delta(\mathbf{r})} = \sum_{i=0,1,2,3} d^{i}(\mathbf{r})  \sigma_i i \sigma_2
    = \sum_{i=0,1,2,3} [d_{\text{ext }}^i\left(\mathbf{r}\right) 
 -\lambda^{i}\left(\mathbf{r}\right) \chi^i\left(\mathbf{r}\right) ]  \sigma_i i \sigma_2
\ee

\section{SIESTA-BdG simulations in practice }\label{subsec:siestabdginpractice}

\subsection{Convergence of \textsc{SIESTA}-BdG simulations}\label{sec:bdg-convergence}

\begin{figure}
    \centering
    \includegraphics[width=\linewidth]{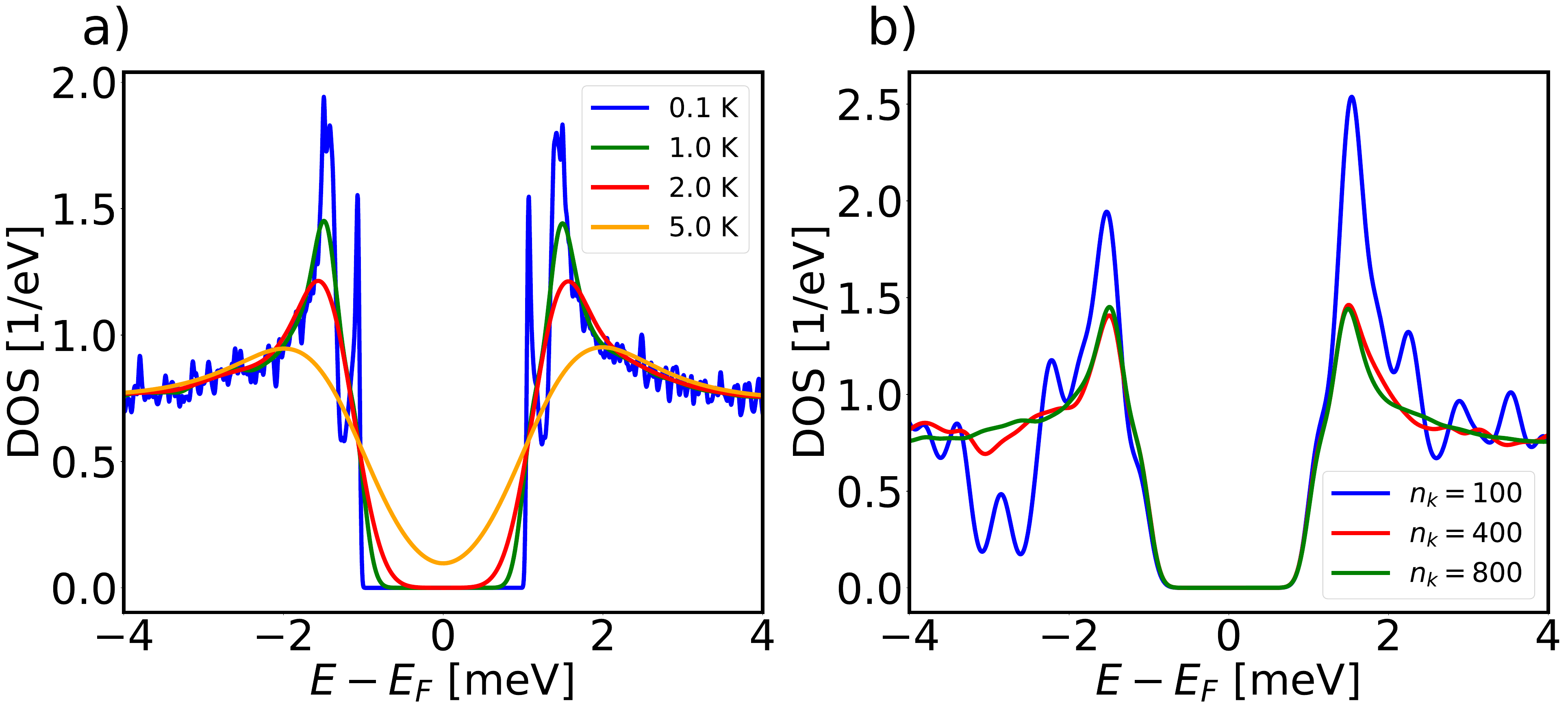}
    \caption{Convergence of \textsc{SIESTA}-BdG calculations for bulk Niobium. 
   a) Dependence of the DOS on smearing temperature at fixed $800\times800\times800$ $\mathbf{k}$-grid.
    Gap anisotropy is only resolved when the smearing temperature is on the order of $0.1$ K or lower.
    b) DOS dependence on $\mathbf{k}$-grid  size ($n_k \times n_k \times n_k$) with a smearing temperature $T=0.1$ K: 
    an extremely dense sampling ($n_k= 800$) is required to converge the tails of the gap and resolve the coherence peaks.
    }
    \label{fig:convergence-tests}
\end{figure}

A rigorous convergence study of simulation parameters is required for the \textsc{SIESTA}-BdG method. This concerns the real-space grid cutoff, the SCF cycle tolerances, the $\mathbf{k}-$grid for the SCF cycle and DOS/PDOS, and the smearing of electron/hole occupations.

The cutoff value of the real-space grid is typically the same order of magnitude as for normal-state \textsc{SIESTA} simulations, as it mostly depend on the choice of the pseudopotentials and on the desired accuracy of the representation of physical quantities (density, potential, charge) on the \textsc{SIESTA} real-space grid.

SCF tolerances  for \textit{non SCF-BdG} and \textit{fixed-$\Delta$}  are similar to   fully-relativistic normal state calculations, namely $10^{-5}$ for the density matrix and  $10^{-4}$ eV for the Hamiltonian). \textit{Full SCF-BdG} calculations require stricter SCF tolerance compared to normal state calculations by one to two orders of magnitude.

The $\mathbf{k}$-grid for the SCF cycles can be converged in steps: first, for the normal metal phase, then the superconducting-state DOS is converged (using the normal-state SCF $\mathbf{k}$-grid), and finally the SCF $\mathbf{k}$-grid is further refined against the shape and size of the superconducting gap.

The smearing temperature and DOS/PDOS sampling of the Brillouin zone need to be pushed beyond the usual converged values for normal-state \textsc{SIESTA} simulations. 
The smearing temperature is related to the Fermi-Dirac occupation function for electrons and holes, as usually defined in first-principles DFT simulations. The smaller the smearing temperature, the higher the energy resolution in the DOS.
The DOS can vary greatly with the smearing temperature, as illustrated in Figure~\ref{fig:convergence-tests}.a. Therefore, the DOS is a good quantity to be used for converging the smearing temperature. As a rule of thumb, setting the temperature to about one-tenth of the superconducting gap usually ensures accurate simulations.

In order to obtain smooth and converged DOS/PDOS it is necessary to use hundreds to thousands of $\mathbf{k}$-points in each direction.
An under-converged sampling results in noisy DOS, which manifests most clearly in the tails that border the superconducting gap, as illustrated in Figure~\ref{fig:convergence-tests}.b for the DOS of bulk Nb computed with different $\mathbf{k}$-grids. 
Performing \textsc{SIESTA}-BdG  DOS/PDOS simulations at such extremely dense sampling of the Brillouin Zone requires the use of the strategies for efficient sampling described in Section~\ref{subsec:stratsample}.

We extended these two strategies to also work for the \textsc{SIESTA}-BdG calculations with Nambu spinors, structuring our algorithm analogously to the one already present in \textsc{SIESTA}~\cite{garcia2020siesta}.
\begin{figure}
    \centering
    \scalebox{1}{
    \begin{tikzpicture}[font=\small,align=center,node distance=0.75cm]

        \node (iter0) at (0,0)
            {\includegraphics[width=0.25\textwidth]{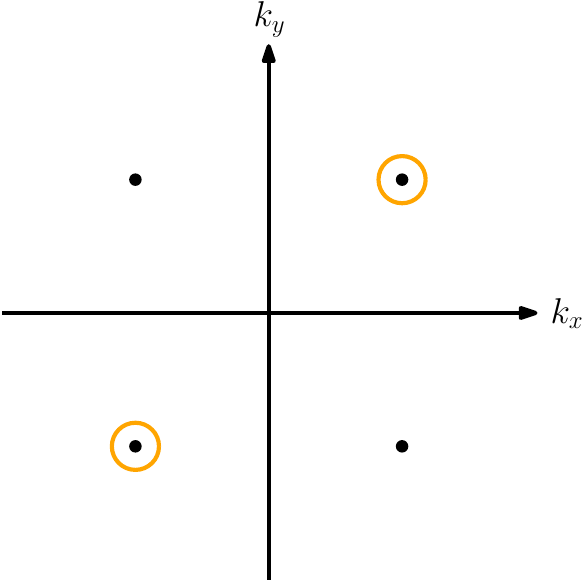}};
        \node (iter0label) at (0,2.5) {Uniform grid};
        \node (iter1) at (5.5,0)
            {\includegraphics[width=0.25\textwidth]{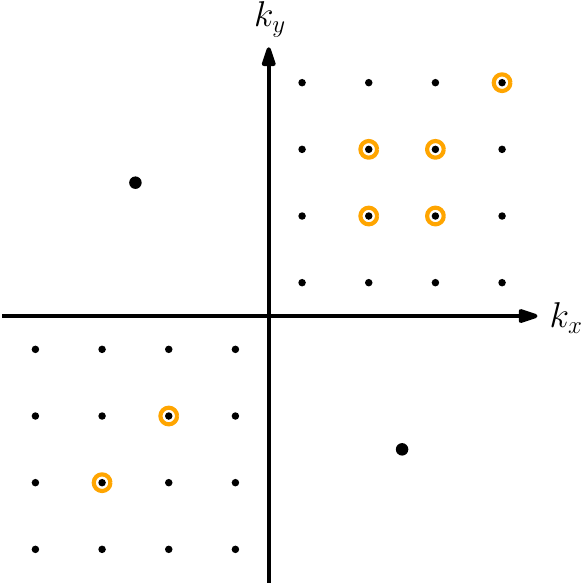}};
        \node (iter1label) at (5.5,2.5) {Uniform grid + 1 zoom};
        \node (iter2) at (11.0,0)
            {\includegraphics[width=0.25\textwidth]{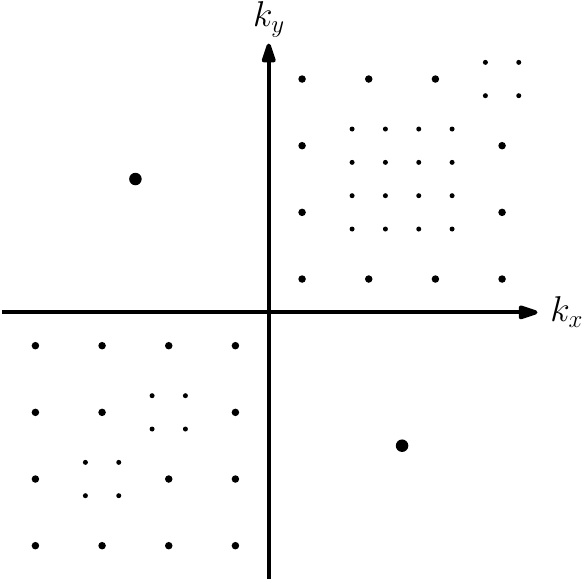}};
        \node (iter2label) at (11.0,2.5) {Uniform grid + 2 zoom};

        \draw[->, >=Straight Barb, line width=8pt] (iter0) -- (iter1) node[pos=0.4,text=white,inner sep=1mm]{zoom};
        \draw[->, >=Straight Barb, line width=8pt] (iter1) -- (iter2) node[pos=0.4,text=white,inner sep=1mm]{zoom};
        
    \end{tikzpicture}}
    \caption{Schematic illustration of the adaptive Brillouin Zone sampling scheme. Starting with a coarse grid, the algorithm determines the points that contribute to the DOS in the region of interest (points circled in orange). To enhance the sampling, the contributing points are replaced by a denser grid (we "zoom" in on the interesting region). This procedure can be repeated to iteratively improve the sampling.}
    \label{fig:adaptiveBZ}
\end{figure}

\subsection{Strategies for efficient sampling of the Brillouin Zone.}\label{subsec:stratsample}
The parallelization techniques allow one to deal with numerous $\mathbf{k}$-points by distributing the work among more processors. Another approach is reducing the number of $\mathbf{k}$-points needed.
For normal state calculations, the tetrahedron method has been shown to be effective in accurately calculating integrals over the Brillouin Zone with relatively few  points~\cite{blochl_improved_1994}. However,
in a typical \textsc{SIESTA}-BdG calculation, the region of interest is limited to a few meV around the Fermi energy. This can be exploited by using an adaptive sampling scheme where a dense sampling grid is only used in the regions of the Brillouin Zone close to the Fermi surface and a coarse grid everywhere else. We implemented this as an iterative procedure, schematically illustrated in Figure~\ref{fig:adaptiveBZ}, using the following steps:

\begin{enumerate}
    \item diagonalize the BdG Hamiltonian $\mathbf{H}^{k}$ and find all the eigenvalues using an initial uniform grid in $\bf{k}$-space,
    \item \label{enum:adaptiveBZstep1} identify which points have at least one eigenvalue that lies in a user-specified energy range,
    \item replace all these points by denser sub-grids centred on the points,
    \item \label{enum:adaptiveBZstep2} diagonalize the Hamiltonian and find the eigenvalues using these new points,
    \item repeat steps~\ref{enum:adaptiveBZstep1} to~\ref{enum:adaptiveBZstep2} until the desired accuracy in the density of states is reached.
\end{enumerate}

\noindent

\begin{figure}
    \centering
    \includegraphics[width=0.6\linewidth]{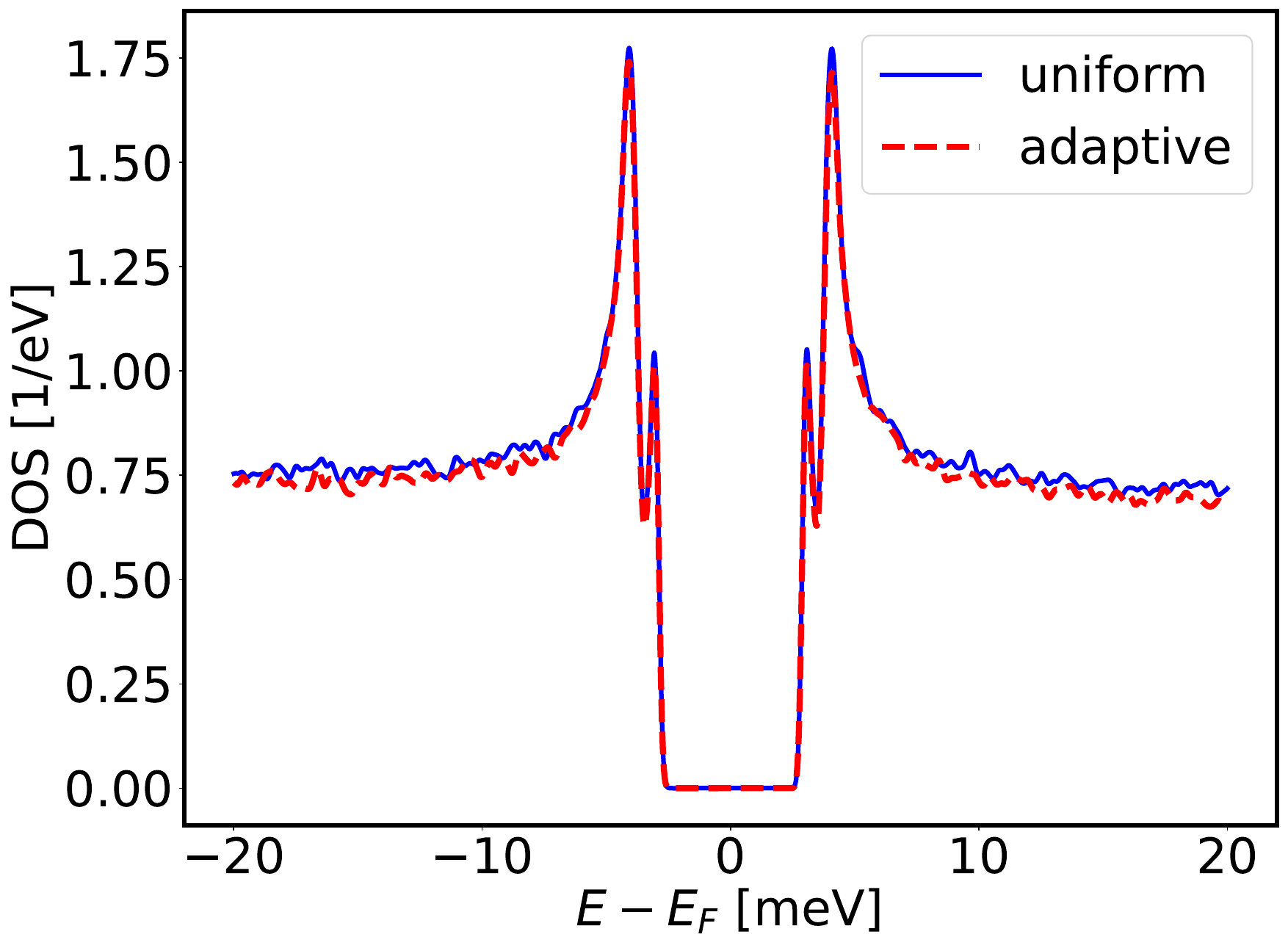}    
    \caption{The DOS of bulk Niobium computed with a uniform $800\times800\times800$ $\mathbf{k}$-grid and an adaptive grid. $\Delta$ has been set to 5 meV. The adaptive grid was generated starting from a $100\times100\times100$ uniform grid and using, for three iterations, a $2\times2\times2$ sub-grid}
    \label{fig:DOS-adaptive-kgrid}
\end{figure}

As an illustration of the effectiveness of this procedure, we compute the DOS of bulk Nb using the adaptive sampling and compare with the DOS computed with a uniform $800\times800\times800$ $k$-grid (Figure~\ref{fig:convergence-tests}.b). The results can be seen in Figure~\ref{fig:DOS-adaptive-kgrid}. The adaptive procedure starts with a $100\times100\times100$ grid and at each iteration (repeated three times) contributing points were replaced 
by a $2\times2\times2$ sub-grid. This procedure allows one to effectively reach the uniform $800\times800\times800$ sampling but only in a small energy window around the Fermi energy.
The adaptive grid reproduces the smooth DOS at a fraction of the computational cost.

\subsection{Limit of applicability of non-SCF BdG solution method}\label{subsec:comparingnonscfbdgfixeddelta}

\begin{figure}
    \centering
    \includegraphics[width=0.45\linewidth]{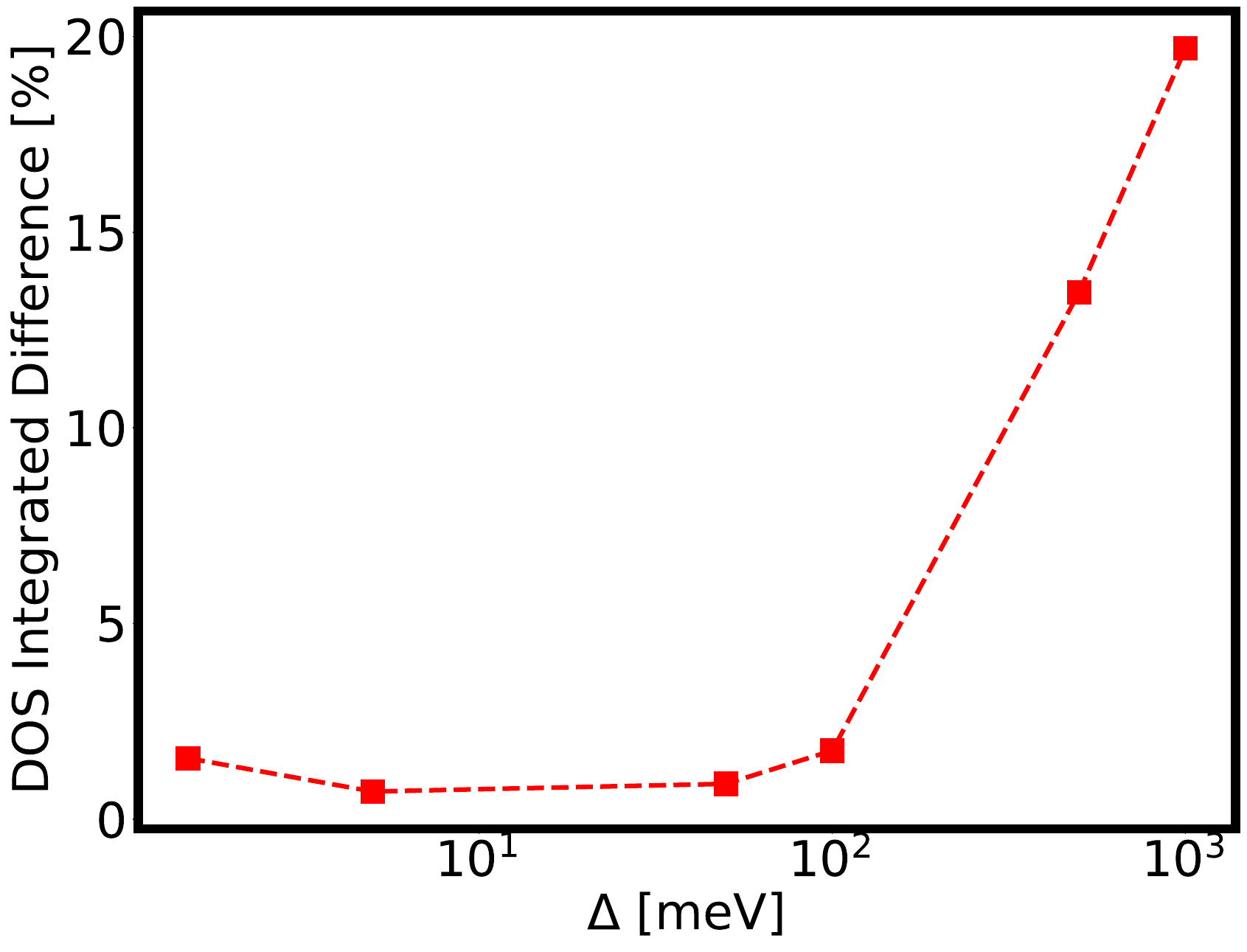}
    \caption{The relative integrated difference in the DOS of bulk Niobium for \textit{non SCF-BdG} and \textit{fixed-$\Delta$} calculations with different magnitudes of the pairing potential $\Delta$.}
    \label{fig:model-rel-int-diff}
\end{figure}

To investigate the differences in results between using the \textit{non SCF-BdG} and \textit{fixed-$\Delta$} approaches, the density of states (DOS) of bulk Niobium was computed for different magnitudes of the pairing potential $\Delta$ using both methods. For each magnitude a real-space initialization of touching hard spheres was used. The results of the two approaches were compared by integrating the absolute differences between the curves. This difference was divided by the integral of the DOS of fixed-$\Delta$ over the same energy range to get a relative integrated difference. The results can be seen in Figure~\ref{fig:model-rel-int-diff}. The differences in the DOS remain small (less than 2\%) up to a $\Delta$ of 100 meV. This is consistent with earlier statements that the superconducting pairing potential is a small perturbation. For large values of $\Delta$ ($\geq1$eV) the difference become significant. This however lies beyond the reasonable strength for Niobium and most known superconductors. Nevertheless, we stress that one should verify which of the approaches is sufficient for the system under study.

\subsection{Comparing SCF-BdG solution methods and real space initializations}\label{sec:bulkPb}
In this Section, we use bulk Pb to compare the three different \textsc{SIESTA}-BdG SCF solution methods introduced in Section~\ref{subsec:solmethod}.
We work in the \textit{superconducting strength representation} of Section~\ref{subsec:specifypairingpot}. 
For the \textit{full SCF-BdG} method, we also compare 
different ways of specifying the superconducting coupling $\lambda(\mathbf{r})$. 

Bulk Pb is a conventional $s$-wave superconductor, where the pairing potential arises from electron-phonon interactions~\cite{BlackfordTunnelingInvestigation1969, LykkenMeasurementSuperconducting1971, HeidEffectSpinorbit2010, SklyadnevaElectronphononInteraction2012, SirohiTransportSpectroscopy2016}. The superconducting gap of lead is U-shaped, the density of states inside the superconducting gap is flat and rises sharply at the edges of the gap (coherence peaks) ~\cite{LykkenMeasurementSuperconducting1971}. The size of the superconducting gap has been measured between 2.3 and 2.5~meV, depending on the crystal orientation~\cite{BlackfordTunnelingInvestigation1969, LykkenMeasurementSuperconducting1971}.

Based on this knowledge, we use a singlet-pairing potential with a strength of $\bar\Delta=$1.25~meV (half of the gap size).
We assume this potential to be uniform and isotropic in real space, i.e., in matrix form, the pairing potential can be written as
\begin{equation}
    \Delta_{\mu\nu}(\mathbf{R}) =
    \bar\Delta
    \left(\begin{array}{cc}
      0 & S_{\mu\nu}(\mathbf{R}) \\
     -S_{\mu\nu}(\mathbf{R}) & 0
    \end{array}\right)
\end{equation}
where $S_{\mu\nu}(\mathbf{R})$ is the overlap matrix. 

\begin{figure}
  \centering
  \includegraphics[width=0.98\textwidth]{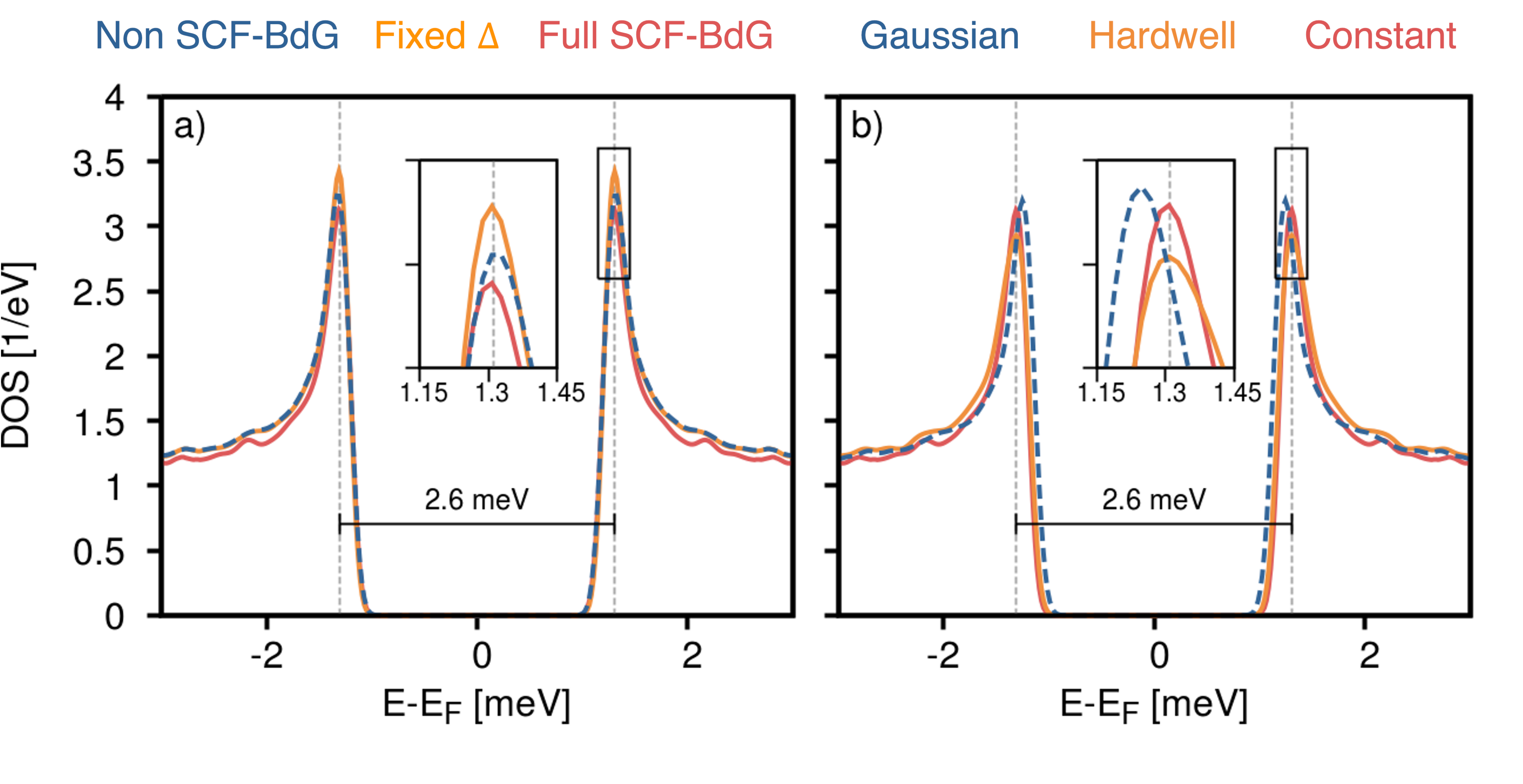}
    \caption{DOS of bulk Pb calculated with the \textsc{SIESTA}-BdG method with smearing temperature $0.01$~meV. (a) Comparison of solution methods \textit{non SCF-BdG}, \textit{fixed-$\Delta$}, and \textit{full SCF-BdG}, assuming isotropic pairing $\Delta(\mathbf{r})=\bar\Delta=1.25$~meV; $\lambda(\mathbf{r})=\bar\lambda=0.315$~eV$\cdot V_{UC}$. 
    (b) Comparison between real-space initialization methods within the \textit{full SCF-BdG} approach 
    (constant, touching spherical hardwells, atom-centered Gaussian ($\sigma=0.3$~\AA)).}
     \label{fig:dos_Pb}
\end{figure}

\begin{figure}
  \centering
  \includegraphics[width=0.54\textwidth]{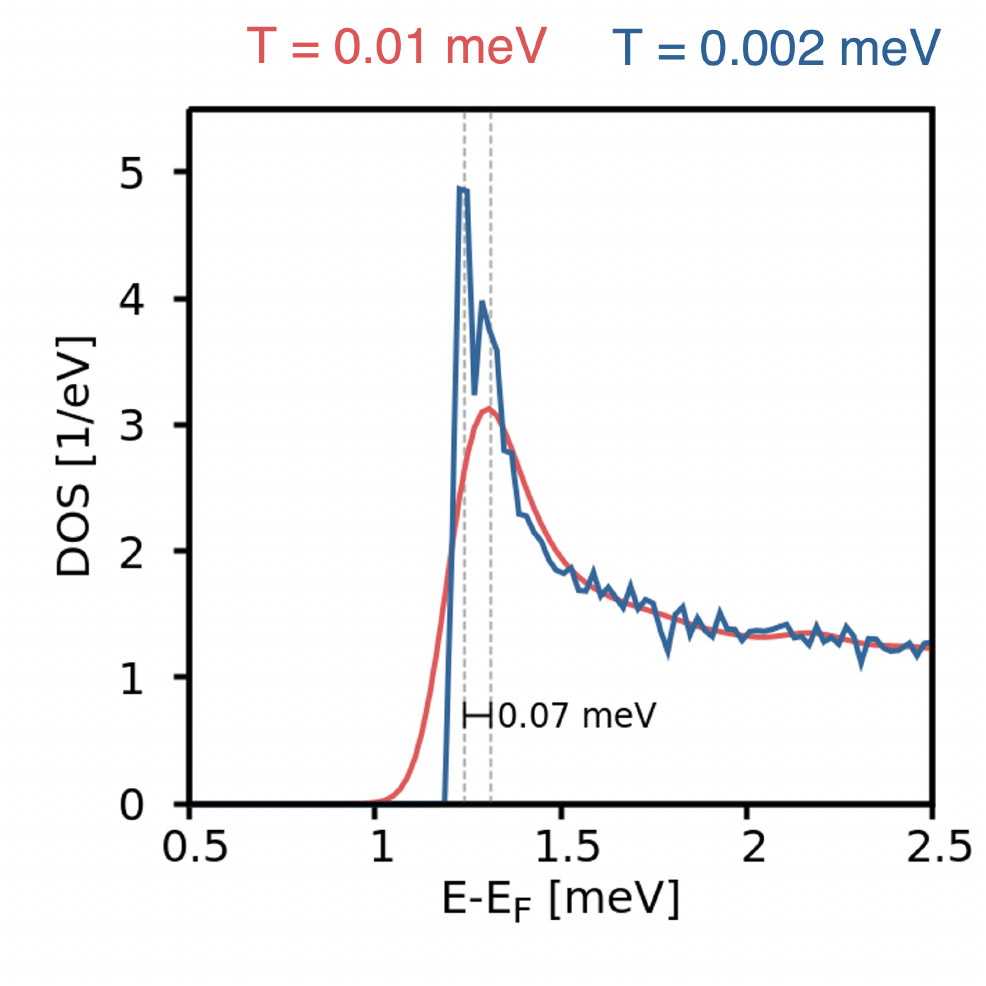}
    \caption{DOS of bulk Pb calculated with \textit{full SCF-BdG}
    with smearing temperature $0.01$~meV (red) and $0.002$~meV (blue) using the constant real-space initialization method as in Figure~\ref{fig:dos_Pb}.c. The plot shows that the computed coherence peak splitting ($0.07$ meV) agrees well with the experimental one $0.1$ meV.}
     \label{fig:dos_Pb_resolved}
\end{figure}

We perform calculations with all three solution methods (\textit{non SCF-BdG}, \textit{fixed-$\Delta$}, and \textit{full SCF-BdG}) to obtain the DOS for the superconducting phase of Pb. 
For the \textit{full SCF-BdG} method in Figure~\ref{fig:dos_Pb}.a, we 
define $\lambda$ isotropic in space and tune its strength to obtain a gap with the same width as for the two other methods finding $\bar{\lambda}=0.315$~eV~$\cdot V_{UC}$, where $V_{UC}$ is the unit cell volume.
The resulting superconducting gap confirms the experiment in shape and magnitude
(Figure~\ref{fig:dos_Pb}.a) irrespective of the solution method used. The coherence peaks at the edge of the gap are sharp, and the density of states forms a small plateau between these peaks. 
Simulations with a smearing temperatures of $0.002$~meV allow us to predict a splitting of the coherence peak of $0.07$ meV, which agrees well with the experimentally observed one ($0.1$ meV) (Figure~\ref{fig:dos_Pb_resolved}).

\begin{figure}
    \includegraphics[width=0.98\textwidth]{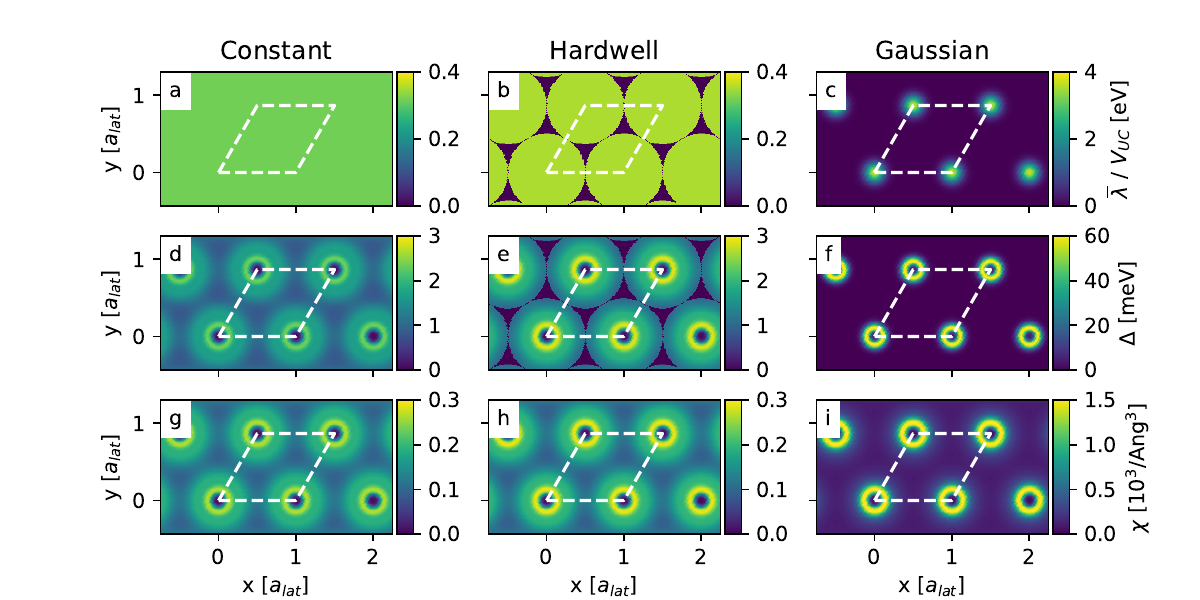}
    \caption{
    Real-space visualization of (a -- c) the superconducting coupling parameter
    $\lambda(\mathbf{r})$, (d -- f) the pairing potential  $\Delta(\mathbf{r})$, and (g -- i) the anomalous density $\chi(\mathbf{r})$ for bulk Pb. These quantities are computed for different initializations of $\lambda(\mathbf{r})$ in the \textit{full SCF-BdG} solution method. The columns correspond to different initialization: (a) constant, (b) touching spherical hardwell ($R = 1.704$~\AA), and (c) Gaussian shape ($\sigma=0.3$~\AA)).
    The intensity of each quantity is represented as a colour gradient.
    The dashed parallelogram indicates the unit cell of Pb with atoms centred on the corners.}
    \label{fig:dos_grid}
\end{figure}

To assess the influence of the real-space initialization of the superconducting coupling $\lambda\mathbf{r})$, we calculate the pairing potential $\Delta(\mathbf{r})$ and the anomalous density $\chi(\mathbf{r})$ of bulk lead using the \textit{full SCF-BdG} approach and the following initializations: 
(i) \textit{constant} $\lambda(\mathbf{r})=\bar\lambda$, 
(ii) touching \textit{spherical hardwell} $\lambda(\mathbf{r})=\bar\lambda$ if $|\mathbf{r}-\mathbf{r}_{\mathrm{Pb}}| < 1.704$~\AA, and 
(iii) \textit{Gaussian} shape $\lambda(\mathbf{r})=\bar\lambda e^{-|\mathbf{r}-\mathbf{r}_{\mathrm{Pb}}|^2/(2\sigma^2)}$; $\sigma=0.3$~\AA~, as illustrated in Figure \ref{fig:dos_grid}. 
In each case, we tune the parameter $\bar\lambda$ to obtain a gap with approximately $2.56$~meV width (Figure \ref{fig:dos_Pb}.b), resulting in $\bar\lambda/V_{UC}=$0.315~eV, 0.350~eV, and 3.69~eV for (i), (ii), and (iii), respectively.

Despite the differences in the functional form of the superconducting coupling, the computed physical properties (DOS and anomalous charge) are qualitatively similar, 
indicating that the physically relevant information is included in the overlap matrix, namely in the projection of $\lambda$ and/or $\Delta$ on the orbitals close to the Fermi level.
The width of the superconducting gap is the same by construction.
The extension of the anomalous charge and the spatial representation of $\Delta$, instead, depend on the choice of $\bar\lambda$  and the specific radius of the modulation function.
The case that differ most is the Gaussian initialization, as those functions are more sharply localized on the atoms.

\subsection{Computational Details}\label{subsec:compdetails}
\textbf{\textsc{SIESTA} and \textsc{SIESTA}-BdG:} The simulations for bulk Nb, Pb, and FeSe were performed with PBE~\cite{PerdewGeneralizedGradient1996} functional,  norm-conserving pseudopotentials from the Pseudo-Dojo database~\cite{vanSettenPseudoDojoTraining2018}, and the default SIESTA double-$\zeta$ polarized (DZP) basis set~\cite{SolerSIESTAMethod2002}.
The crystal structure of each system was relaxed with a maximum stress tolerance of $0.01$ GPa.

For bulk Nb, the KS equations were solved using $T = 0.1$ K, a $120\times120\times120$ $\mathbf{k}$-grid, and a mesh cutoff of $600$ Ry. Structural relaxation was performed with $T = 1$ meV , and $40\times40\times40$  $\mathbf{k}$-grid.
The DOS was sampled on a $400\times400\times400$ $\mathbf{k}$-grid and then refined iterating the Adaptive Brillouin Zone sampling three times with a $2\times2\times2$ sub-grid.
A spherical square well potential with a constant value of $\lambda = 0.188$~eV$\cdot V_\mathrm{sphere}$ inside the sphere was used (radius $r = 1.4136$ \AA) to solve the SIESTA-BdG equations.

For bulk Pb, the KS equations were solved using $T = 1$K, a $60\times60\times60$ $\mathbf{k}$-grid, and a mesh cutoff of $2000$ Ry. The DOS was sampled on a $300\times300\times300$ $\mathbf{k}$-grid and then refined with a $3\times3\times3$ sub-grid.

The FeSe simulations are converged for a smearing temperature of 0.5~meV, a $29\times29\times29$ \textbf{k}-grid for KS-equations, and a mesh cutoff of 1300~Ry. The DOS was sampled on a 480\ttimes{}480\ttimes{}120 \textbf{k}-grid.

\textbf{Quantum transport:} We solve the KS equations using a smearing temperature of 1~meV, a  1\ttimes{}1\ttimes{}151 \textbf{k}-grid, and a mesh cutoff of 800~Ry. The simulations are performed with the LDA functional as parameterized by Perdew and Zunger~\cite{PerdewSelfinteractionCorrection1981} and norm-conserving fully relativistic pseudopotentials from the Pseudo-Dojo database~\cite{vanSettenPseudoDojoTraining2018}.
The atomic positions were calculated according to the structure of an ideal nanotube with lattice constant 2.46~\AA. 
We perform a \textsc{SIESTA}-BdG calculation using the \textit{non SCF-BdG} method with an pairing of 5~meV in the superconducting section of the scattering region and the right electrode, and set the super conducting pairing to 0 in the normal conductor section. While it is well known that the superconducting order parameter decays as a power law in low-dimensional nanostructures~\cite{GonzalezCooperpairPropagation2007, GonzalezCriticalCurrents2008}, we enforce a sharp decay here for a proof of concept.
In order to calculate the quantum conductance, we use a custom Python script (Supplementary Material).

\section{Scattering theory for quantum transport with the \textsc{SIESTA}-BdG formalism}\label{app:transport}

The relationship between current and voltage in superconductors is fundamentally different from that encountered in normal conductors. 
In a non-superconducting device, the current through any electrode is given by
\begin{align}
     I_\mathfrak{e} = \int_{-\infty}^\infty\!\!d\varepsilon
        [M_\mathfrak{e}(\varepsilon) - R_\mathfrak{e}(\varepsilon)] f_\mathfrak{e}(\varepsilon) 
        - \sum_{\mathfrak{e}'\neq\mathfrak{e}} T_{\mathfrak{e}\mathfrak{e}'}(\varepsilon) f_{\mathfrak{e}'}(\varepsilon)
\end{align}
where $\mathbf{M}(\varepsilon)$ is the number of channels in the electrode $\mathfrak{e}$ at energy $\varepsilon$, $\mathbf{R}(\varepsilon)$ is the reflection function, $\mathbf{T}_{\mathfrak{e},\mathfrak{e}'}(\varepsilon)$ is the transmission function between two electrodes $\mathfrak{e}$ and $\mathfrak{e}'$, and $f_{\mathfrak{e}}(\varepsilon)$ is the occupation function for the electrode $\mathfrak{e}$. Each electrode has its own chemical potential ($\mu_\mathfrak{e}$) and temperature ($T_\mathfrak{e}$).
The $\mathbf{k}$ dependence of all quantities here is implicit.

The transmission and reflection functions are directly related to the scattering matrix $\textbf{s}_{\mathfrak{e}\mathfrak{e}'}(\varepsilon)$:
\begin{align}
    T_{\mathfrak{e}\mathfrak{e}'}(\varepsilon) &= \mathrm{Tr}\left\{
        \textbf{s}_{\mathfrak{e}\mathfrak{e}'}^\dagger(\varepsilon)
        \textbf{s}_{\mathfrak{e}\mathfrak{e}'}(\varepsilon)
    \right\}&
    R_\mathfrak{e}(\varepsilon) &= M_\mathfrak{e}(\varepsilon) - \mathrm{Tr}\left\{
        \textbf{s}_{\mathfrak{e}\mathfrak{e}}^\dagger(\varepsilon)
        \textbf{s}_{\mathfrak{e}\mathfrak{e}}(\varepsilon)
    \right\}
\end{align}
Furthermore, the unitarity of the scattering matrix implies $M_\mathfrak{e} = R_\mathfrak{e} + \sum_{\mathfrak{e}'\neq\mathfrak{e}}T_{\mathfrak{e}\mathfrak{e}'}$ \cite{DattaElectronicTransport1995}. This relationship allows us to rewrite the current as a sum of pairwise currents.
\begin{align}
    I_\mathfrak{e} &= \sum_{\mathfrak{e}'\neq\mathfrak{e}} I_{\mathfrak{e},\mathfrak{e}'} &
    I_{\mathfrak{e},\mathfrak{e}'} &= \int_{-\infty}^\infty\!\!d\varepsilon \, T_{\mathfrak{e}\mathfrak{e}'}(\varepsilon)\left[f_\mathfrak{e}(\varepsilon) - f_{\mathfrak{e}'}(\varepsilon)\right]
\end{align}
In the low-bias limit the difference in the Fermi-occupation function of the electrodes is proportional to the difference in their chemical potential and the pairwise currents my be written as.
\begin{align}
    I_{\mathfrak{e},\mathfrak{e}'} &\approx T_{\mathfrak{e}\mathfrak{e}'}(\mu) (\mu_{\mathfrak{e}'}-\mu_\mathfrak{e})
\end{align}
In other words, the current is proportional to the applied bias.

In the case of superconducting electrodes or devices, the transmission and reflection probabilities become 2\ttimes2 matrices (Eq.~\ref{eq:RTscmatrices})~\cite{LambertGeneralizedLandauer1991}.

We obtain the reflection and transmission probabilities using Green's function techniques, adopting an approach similar to the generalized Fisher-Lee relation~\cite{boumrar2020equivalence}.
\begin{align}
    T_{\mathfrak{e}\mathfrak{e}'}^{\alpha\beta}(\varepsilon) &= 
        \mathrm{Tr}\left\{\mathbf s^{\alpha\beta}_{\mathfrak{e}\mathfrak{e}'}(\varepsilon) \mathbf s^{\alpha\beta}_{\mathfrak{e}\mathfrak{e}'}(\varepsilon)^\dagger\right\}\\
    R_{\mathfrak{e}\mathfrak{e}'}^{\alpha\beta}(\varepsilon) &= 
        \mathrm{Tr}\left\{\mathbf s^{\alpha\beta}_{\mathfrak{e}\mathfrak{e}'}(\varepsilon) \mathbf s^{\alpha\beta}_{\mathfrak{e}\mathfrak{e}'}(\varepsilon)^\dagger\right\}\\
    s^{\alpha\beta}_{\mathfrak{e}\mathfrak{e}'} &= -\delta_{\alpha\beta} \delta_{\mathfrak{ee}'} M^\alpha_\mathfrak{e}(\varepsilon) + i \mathbf \Gamma_\mathfrak{e}^{1/2}(\varepsilon) \mathbf G(\varepsilon+i0^+) \mathbf \Gamma_{\mathfrak{e}'}^{1/2}(\varepsilon)\\
    \mathbf G(\varepsilon+i0^+) &=(\varepsilon+i0^+) \mathbf S  - \mathbf H - \sum_\mathfrak{e} \mathbf\Sigma_\mathfrak{e}(\varepsilon)
\end{align}
where $\Gamma_\mathfrak{e}(\varepsilon) = i \left[\mathbf\Sigma_\mathfrak{e}(\varepsilon) - \mathbf\Sigma^\dagger_\mathfrak{e}(\varepsilon)\right]$ is the electrode broadening matrix, $\mathbf\Sigma_\mathfrak{e}(\varepsilon)$ is the electrode self-energy, $\mathbf G(\varepsilon+i0^+)$ is the retarded Green's function of the open quantum system, and $M^\alpha_\mathfrak{e}(\varepsilon)$ is the number of electrode channels. 

The electrode self-energies are calculated using the Lopez-Sancho/Lopez-Sancho algorithm~\cite{sancho1984quick}. The number of channels $M^\alpha_\mathfrak{e}(\varepsilon)$ is determined in a separate calculation of the bulk electrodes.

These additional scattering probabilities have to be taken into account when calculating the current through any electrode. Taking into account the opposite charge of electrons and holes, the current through the electrode $\mathfrak{e}$ is given by:
\begin{align}
    I_{\mathfrak{e}} = \frac{e}{h}
        \int_{-\infty}^\infty\Big\{
        &\underbrace{
            \left[
                M_\mathfrak{e}(\varepsilon) - R^{pp}_\mathfrak{e}(\varepsilon) + R^{hp}_\mathfrak{e}(\varepsilon)
            \right] {f}_\mathfrak{e}(\varepsilon) 
            - \sum_{\mathfrak{e}'\neq\mathfrak{e}} \left[
                T^{pp}_{\mathfrak{e}\mathfrak{e}'}(\varepsilon) - 
                T^{hp}_{\mathfrak{e}\mathfrak{e}'}(\varepsilon)
            \right] {f}_{\mathfrak{e}'}(\varepsilon)}_{\text{electronic contribution}}
            \nonumber\\            
            -&\underbrace{\left[
                M_\mathfrak{e}(\varepsilon) - R^{hh}_\mathfrak{e}(\varepsilon) + R^{ph}_\mathfrak{e}(\varepsilon)
            \right] {\bar{f}}_\mathfrak{e}(\varepsilon) 
            - \sum_{\mathfrak{e}'\neq\mathfrak{e}} \left[
                -T^{hh}_{\mathfrak{e}\mathfrak{e}'}(\varepsilon)  
                +T^{ph}_{\mathfrak{e}\mathfrak{e}'}(\varepsilon) ]
            \right] {\bar{f}}_{\mathfrak{e}'}(\varepsilon)}_{\text{hole contribution}}
        \Big\}d\varepsilon.
\end{align}
\begin{align}
    I_{\mathfrak{e}} = \frac{e}{h}
        \int_{-\infty}^\infty\Big\{
        &\underbrace{
            \left[
                M_\mathfrak{e}(\varepsilon) - R^{pp}_\mathfrak{e}(\varepsilon) + R^{hp}_\mathfrak{e}(\varepsilon)
            \right] {f}_\mathfrak{e}(\varepsilon) 
            - \sum_{\mathfrak{e}'\neq\mathfrak{e}} \left[
                T^{pp}_{\mathfrak{e}\mathfrak{e}'}(\varepsilon) - 
                T^{hp}_{\mathfrak{e}\mathfrak{e}'}(\varepsilon)
            \right] {f}_{\mathfrak{e}'}(\varepsilon)}_{\text{electronic contribution}}
            \nonumber\\            
            -&\underbrace{\left[
                M_\mathfrak{e}(\varepsilon) - R^{hh}_\mathfrak{e}(\varepsilon) + R^{ph}_\mathfrak{e}(\varepsilon)
            \right] {\bar{f}}_\mathfrak{e}(\varepsilon) 
            - \sum_{\mathfrak{e}'\neq\mathfrak{e}} \left[
                -T^{hh}_{\mathfrak{e}\mathfrak{e}'}(\varepsilon)  
                +T^{ph}_{\mathfrak{e}\mathfrak{e}'}(\varepsilon) ]
            \right] {\bar{f}}_{\mathfrak{e}'}(\varepsilon)}_{\text{hole contribution}}
        \Big\}d\varepsilon.
\end{align}
Here, $\bar{f}(\varepsilon) = 1 - f(\varepsilon)$ is the occupation function of the hole-like states in the electrode.
As in a normal metal, the unitarity of the scattering matrix links the number of electrode channels to the reflection and transmission functions~\cite{LambertPhasecoherentTransport1998}: 
\begin{align}
    M_\mathfrak{e}(\varepsilon) 
    &= R^{pp}_\mathfrak{e}(\varepsilon) 
    + R^{ph}_\mathfrak{e}(\varepsilon) 
    + \sum_{\mathfrak{e}'\neq\mathfrak{e}} [
        T^{pp}_{\mathfrak{e}\mathfrak{e}'}(\varepsilon) 
        + T^{ph}_{\mathfrak{e}\mathfrak{e}'}(\varepsilon)]
    \nonumber\\ 
    &= R^{hh}_\mathfrak{e}(\varepsilon) 
    + R^{hp}_\mathfrak{e}(\varepsilon) 
    + \sum_{\mathfrak{e}'\neq\mathfrak{e}} [
        T^{hh}_{\mathfrak{e}\mathfrak{e}'}(\varepsilon) 
        + T^{hp}_{\mathfrak{e}\mathfrak{e}'}(\varepsilon)].
\end{align}
Using these identities, we can rewrite the current through electrode $I_e$ (Eq.~\ref{eq:currentsc}) as
\begin{align}
    I_{\mathfrak{e}} = \frac{e}{h}
        \int_{-\infty}^\infty\Big\{
            &[R^{hp}_\mathfrak{e}(\varepsilon) 
            + R^{ph}_\mathfrak{e}(\varepsilon)] [f_{\mathfrak{e}}(\varepsilon) - \bar{f}_{\mathfrak{e}}(\varepsilon)] 
            \nonumber\\
            &+\sum_{\mathfrak{e}'\neq\mathfrak{e}} 
                [ 
                T^{pp}_{\mathfrak{e}\mathfrak{e}'}(\varepsilon)
                + T^{hh}_{\mathfrak{e}\mathfrak{e}'}(\varepsilon)  ][
                    f_{\mathfrak{e}}(\varepsilon) - f_{\mathfrak{e}'}(\varepsilon)
                ]
            \nonumber\\
            &+ \sum_{\mathfrak{e}'\neq\mathfrak{e}} 
                T^{hp}_{\mathfrak{e}\mathfrak{e}'}(\varepsilon) 
                [\bar{f}_{\mathfrak{e}}(\varepsilon) 
                -f_{\mathfrak{e}'}(\varepsilon)]
                +T^{ph}_{\mathfrak{e}\mathfrak{e}'}(\varepsilon) 
                [f_{\mathfrak{e}}(\varepsilon) 
                -\bar{f}_{\mathfrak{e}'}(\varepsilon)]
        \Big\}d\varepsilon
\end{align}
where $\bar{f}_{\mathfrak{e}}(\varepsilon)$ is the occupation of hole states in the electrode $\mathfrak{e}$. Moving to the low-bias limit ($
    f_\mathfrak{e}(\varepsilon) 
    = f(\varepsilon) - \frac{\partial{f}}{\partial{\varepsilon}}(\varepsilon) [\mu_\mathfrak{e}-\mu] 
$), and invoking particle-hole symmetry
allows us to rewrite the expression of the current as (Eq.~\ref{eq:currentsclowbias}):
\begin{align}
    I_{\mathfrak{e}} 
        \approx& \frac{e}{h}\sum_{\mathfrak{e}'\neq\mathfrak{e}}
            T^{pp}_{\mathfrak{e}\mathfrak{e}'}(\mu) [\mu_{\mathfrak{e}'} - \mu_{\mathfrak{e}}]
        +\sum_{\mathfrak{e}'\neq\mathfrak{e}} 
            T^{hp}_{\mathfrak{e}\mathfrak{e}'}(\mu) [\mu_{\mathfrak{e}'} + \mu_{\mathfrak{e}} - 2\mu]
        +2 R^{hp}_\mathfrak{e}(\mu) (\mu_\mathfrak{e}-\mu)
\end{align}

\newpage

\bibliographystyle{unsrt}
\bibliography{bibliography}

\end{document}